\documentclass[onecolumn,titlepage,superscriptaddress,nofootinbib,11pt]{revtex4-1}
\usepackage{amsmath,amsthm,amsfonts,amscd,amssymb} 
\usepackage{braket}
\usepackage{slashed}
\usepackage[hidelinks]{hyperref}
\usepackage{graphicx}
\usepackage{xcolor}

\newcommand{\be}{\begin{equation}}
\newcommand{\ee}{\end{equation}}

\newcommand{\bear}{\begin{eqnarray}}
\newcommand{\eear}{\end{eqnarray}}

\newcommand\snowmass{\begin{center}\rule[-0.2in]{\hsize}{0.01in}\\\rule{\hsize}{0.01in}\\
\vskip 0.1in Submitted to the  Proceedings of the US Community Study\\ 
on the Future of Particle Physics (Snowmass 2021)\\ 
\rule{\hsize}{0.01in}\\\rule[+0.2in]{\hsize}{0.01in} \end{center}}

\begin{document}

\title{Snowmass 2021 White Paper: The Windchime Project}
\author{\textbf{The Windchime Collaboration:} Alaina Attanasio}
\affiliation{Purdue University, West Lafayette, IN, USA}
\author{Sunil A. Bhave}
\affiliation{Purdue University, West Lafayette, IN, USA}
\author{Carlos Blanco}
\affiliation{Princeton University, Princeton, NJ, USA}
\affiliation{Stockholm University, Stockholm, Sweden}
\author{Daniel Carney}
\affiliation{Lawrence Berkeley National Laboratory, Berkeley, CA, USA}
\author{Marcel Demarteau}
\affiliation{Oak Ridge National Laboratory, Oak Ridge, TN, USA}
\author{Bahaa Elshimy}
\affiliation{Purdue University, West Lafayette, IN, USA}
\author{Michael Febbraro}
\affiliation{Oak Ridge National Laboratory, Oak Ridge, TN, USA}
\author{Matthew A. Feldman}
\affiliation{Oak Ridge National Laboratory, Oak Ridge, TN, USA}
\author{Sohitri Ghosh}
\affiliation{ Joint Quantum Institute/Joint Center for Quantum Information and Computer Science, University of Maryland, College Park/National Institute of Standards and Technology, Gaithersburg, MD, USA}
\author{Abby Hickin}
\affiliation{Purdue University, West Lafayette, IN, USA}
\author{Seongjin Hong}
\affiliation{Oak Ridge National Laboratory, Oak Ridge, TN, USA}
\author{Rafael F. Lang}
\affiliation{Purdue University, West Lafayette, IN, USA}
\author{Benjamin Lawrie}
\affiliation{Oak Ridge National Laboratory, Oak Ridge, TN, USA}
\author{Shengchao Li}
\affiliation{Purdue University, West Lafayette, IN, USA}
\author{Zhen Liu}
\affiliation{School of Physics and Astronomy, University of Minnesota, Minneapolis, MN, USA}
\author{Juan P. A. Maldonado}
\affiliation{Purdue University, West Lafayette, IN, USA}
\affiliation{Universidad Nacional de Colombia, Bogotá, Colombia}
\author{Claire Marvinney}
\affiliation{Oak Ridge National Laboratory, Oak Ridge, TN, USA}
\author{Hein Zay Yar Oo}
\affiliation{Purdue University, West Lafayette, IN, USA}
\author{Yun-Yi Pai}
\affiliation{Oak Ridge National Laboratory, Oak Ridge, TN, USA}
\author{Raphael Pooser}
\affiliation{Oak Ridge National Laboratory, Oak Ridge, TN, USA}
\author{Juehang Qin}
\affiliation{Purdue University, West Lafayette, IN, USA}
\author{Tobias J. Sparmann}
\affiliation{Purdue University, West Lafayette, IN, USA}
\author{Jacob M. Taylor}
\affiliation{ Joint Quantum Institute/Joint Center for Quantum Information and Computer Science, University of Maryland, College Park/National Institute of Standards and Technology, Gaithersburg, MD, USA}
\author{Hao Tian}
\affiliation{Purdue University, West Lafayette, IN, USA}
\author{Christopher Tunnell}
\affiliation{Rice University, Houston, TX, USA}

\medskip

\date{\today}

\begin{abstract}
The absence of clear signals from particle dark matter in direct detection experiments motivates new approaches in disparate regions of viable parameter space. In this Snowmass white paper, we outline the Windchime project, a program to build a large array of quantum-enhanced mechanical sensors. The ultimate aim is to build a detector capable of searching for Planck mass-scale dark matter purely through its gravitational coupling to ordinary matter. In the shorter term, we aim to search for a number of other physics targets, especially some ultralight dark matter candidates. Here, we discuss the basic design, open R\&D challenges and opportunities, current experimental efforts, and both short- and long-term physics targets of the Windchime project.
\end{abstract}

\snowmass 

\maketitle

\tableofcontents

\section{Introduction}

All evidence to date for the existence of dark matter (DM) is through its gravitational coupling to visible matter. On the other hand, all direct detection searches for dark matter to-date have necessarily had to assume some additional coupling to the Standard Model, for example a Weak nuclear coupling in the case of WIMPs, or gluonic/photonic couplings in the case of axions. A clearly desirable target would be to search for particle DM directly through its gravitational coupling alone. 

Recently, it has been suggested that a purely gravitational direct detection strategy could be possible, albeit very challenging, with a terrestrial experiment \cite{Hall:2016usm,Kawasaki:2018xak,Carney:2019pza}. This idea leverages the incredible and rapid progress in quantum readout and control of mechanical sensing devices with optical or microwave light \cite{aspelmeyer2014cavity,blencowe2004quantum,SnowMassWP-IFquantum}. These devices have been demonstrated as a promising platform for a searches for number of dark matter candidates \cite{carney2020mechanical}, spanning the ultralight \cite{graham2016dark,Carney:2019cio,Manley:2020mjq,SnowMassWP-CF2ULDM}, light \cite{Afek:2021vjy}, and up to WIMP-scale and heavier mass ranges \cite{Monteiro:2020wcb}. In particular, Ref. \cite{Carney:2019pza} suggested that a large array consisting of at least $10^6$ mechanical sensors, each with mass around the gram scale, could be sensitive to the gravitational signatures of dark matter with masses around the Planck scale $m_{\rm Pl} \approx 2 \times 10^{18}~{\rm GeV} \approx 4~{\mu g}$. See the Snowmass 2021 community whitepaper \cite{SnowMassWP-CF1UHDM} for an overview of these ultraheavy dark matter candidates.

In this Snowmass white paper, we outline a nascent experimental effort, which we dub the Windchime project, to develop such dark matter detectors. The core program is to construct and operate arrays of many quantum-limited mechanical accelerometers in parallel. Such a system would be uniquely capable of searching for a number of interesting signals, with gravitational dark matter detection a very long-term goal. Many technical developments are needed, spanning four key axes: thermal isolation, quantum measurement noise below the standard quantum limit, scaling in the number of sensors and their readout, and data handling and analysis techniques for continuous data streams from many detectors.  Many shorter-term physics opportunities will be enabled during the development of these technologies, and the R\&D program will have a vast number of applications beyond the search for dark matter. We outline the technical challenges, physics opportunities, our current efforts, and pathways to the fulfillment of the long-term program.

\section{Detector concept}
\label{concept}

\begin{figure}[htbp]
\centering
$\vcenter{\hbox{\includegraphics[scale=0.4]{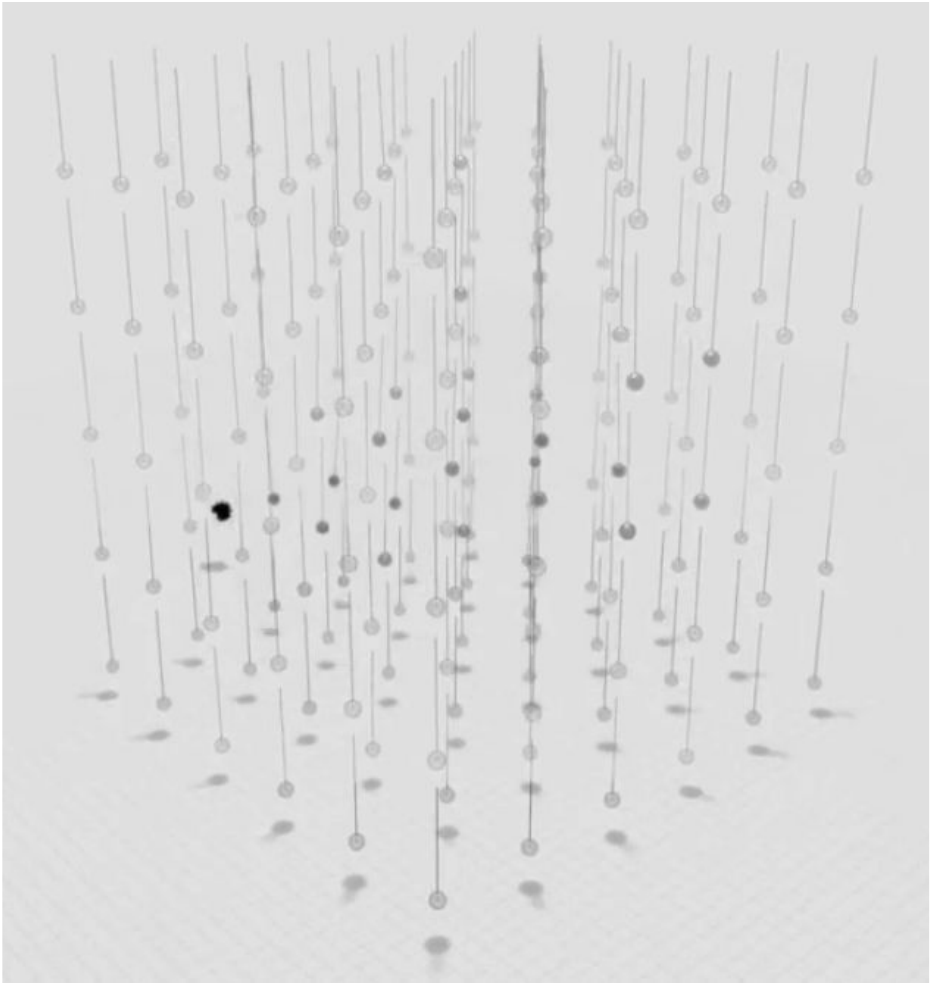}}}~~~~~ \vcenter{\hbox{\includegraphics[scale=0.8]{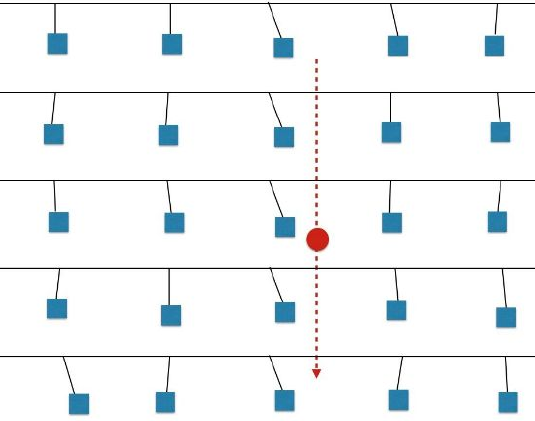}}}~~~~~ \vcenter{\hbox{\includegraphics[scale=0.5]{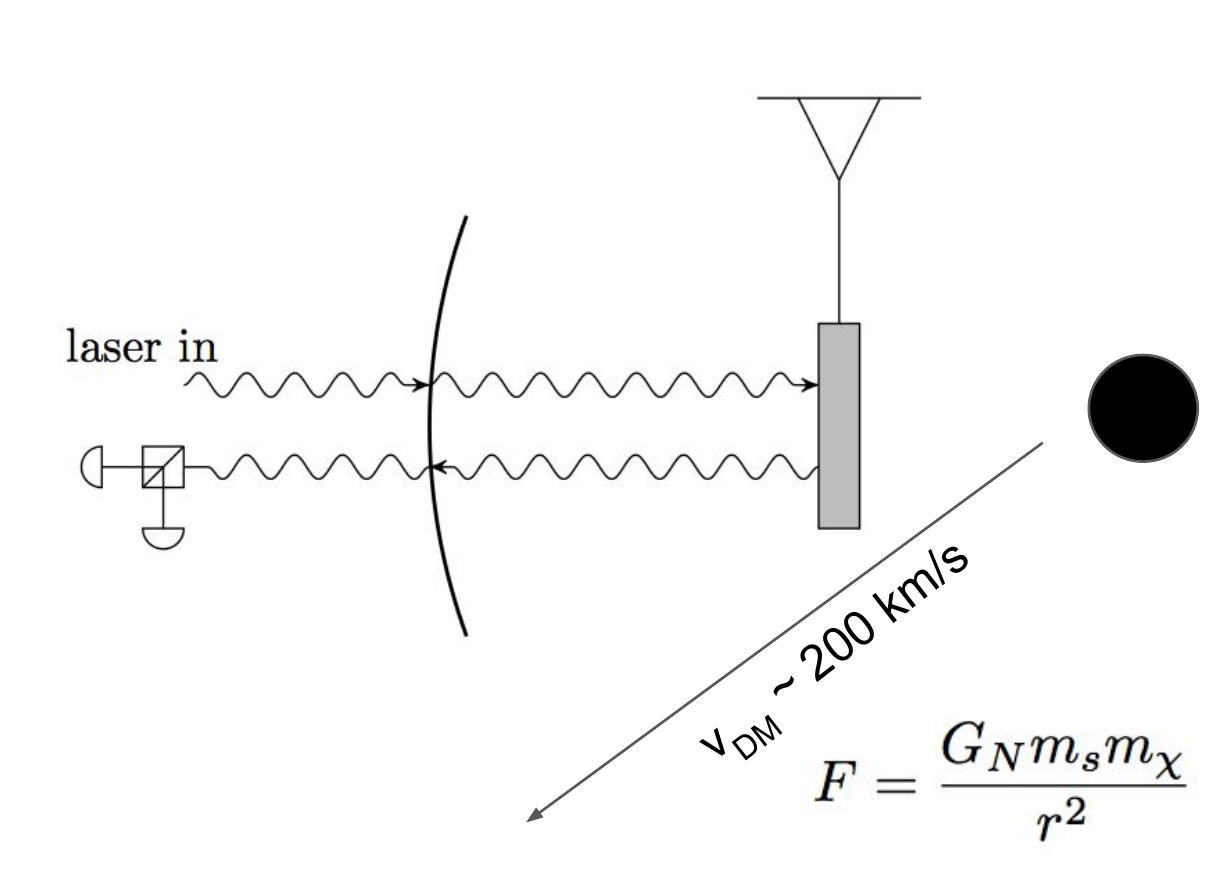}}}$ 
\caption{Schematic illustration of the Windchime detector concept. Left: an array of mechanical sensors, here depicted as suspended pendula, with a potential DM track signal. Center: cross-section, emphasizing the ``track'' signal. Right: single-sensor depiction of a gravitational DM event. Here, for conceptual illustration, the sensor is depicted in a small optical cavity with laser readout. In practice, readout through either fibers or microwave transmission lines will be more convenient for a densely packed array.}
\label{figure-schematic}
\end{figure}

The central concept of the Windchime detector is to use an array of mechanical sensors, operated as a single detector, see Fig.~\ref{figure-schematic} (and Ref.~\cite{Carney:2019pza} for more details). This setup should be uniquely sensitive to signals which are coherent over one, many, or even all of the mechanical devices. Examples include both gravitationally-coupled, localized DM, which acts coherently on sensors near its location (Fig.~\ref{figure-schematic}, center), as well as ultralight DM, which can act coherently on many or all of the sensors (Fig.~\ref{figure-schematic}, right). In this section, we provide a brief, high-level description of both the individual mechanical sensing elements, as well as the operation of a large array of these. More details about physics signals of interest are then given in Sec.~\ref{section-targets}, followed by details about possible physical implementations of the detector concepts in Sec.~\ref{section-RD}.

The individual mechanical sensors of the array are optomechanical/electromechanical systems. The essential architecture is a mechanical element (e.g., suspended reflective pendula \cite{abramovici1992ligo,Catano-Lopez:2019yqb}, optically levitated dielectric spheres \cite{delic2020cooling,Monteiro:2020wcb}, or magnetically levitated superconducting masses \cite{moody2002three,romero2012quantum}), coupled to a readout and control system consisting of an optical or microwave field. The readout field encodes the position or momentum of the mechanical element continuously in time, so the data stream per sensor is a time series $\mathbf{x}(t)$ or $\mathbf{p}(t)$. Using linear response, this can be converted in software to a continuous measurement of an external force $\mathbf{F}(t)$. The core idea is to use this readout to look for the small impulse $\Delta \mathbf{p} = \int \mathbf{F}(t) dt$ imparted to the mechanical system while a DM candidate acts on it. Sensors may be designed with readout on multiple axes, enabling detection of the full \emph{vector} $\Delta \mathbf{p}$, i.e., enabling perfect directional sensitivity.

Placing many of these sensors in an array offers a variety of important advantages. For signals coherent over $N$ sensors, each of which has independent noise, the signal-to-noise (SNR) can be improved by standard $\sqrt{N}$ statistics. Crucially, in the case of ultraheavy DM detected through gravity, the signal is coherent over a ``track'' of sensors, as shown in Fig.~\ref{figure-schematic}. This enables both $\sqrt{N}$ SNR enhancement and, even more importantly, exquisite background rejection. Almost all traditional DM detection backgrounds like cosmic rays or neutrinos will act only on one or a few individual sensors, or be at faster speeds than those expected from DM. Therefore, the correlated track-like signature of a passing heavy DM particle makes a unique signal. This way, not only the direction of the DM itself, but also the speed of a passing DM object, can be determined.

\section{Physics targets}
\label{section-targets}

A full realization of the Windchime detector would have exquisite sensitivity in a number of channels. Any signal that generates a small acceleration or force coherent over scales comparable to the detector will be uniquely accessible. Here, we outline a few targets already identified as interesting examples. Further theory input for potential possibly interesting signals would be of substantial value.

The central, long-term target of this project is Planck-scale DM detection through gravity alone. This target sets the basic design parameters of the experiment. Because the flux of DM is expected to be approximately
\begin{equation}
\Phi_{\chi} = \frac{\rho_{\chi}}{m_{\chi}} v \approx \frac{1}{\rm yr~m^2} \times \frac{m_{\rm Pl}}{m_{\chi}},
\end{equation}
we need a total detector cross-section of at least around a square meter~\cite{Bramante:2018qbc}. Moreover, the sensitivity of an accelerometer scales with its mass like $\sqrt{m}$, while the gravitational signal scales like $1/b$, where $b$ is the typical impact parameter. Numerically, this leads one to look at gram-scale sensors, spaced at approximately centimeter distances, in an array roughly $1000$ sensors on a side. We discuss the substantial technical challenges to reaching this detection target in detail in Section~\ref{section-RD}.

Prior to this long-term gravitational detection target, a number of other DM candidates can be probed. In terms of localized, particle-like excitations \cite{SnowMassWP-CF1UHDM}, this includes composite DM and solitons coupled to the Standard Model (SM) through some new non-gravitational force. As long as this force has a range of order the size of a sensor, the detection problem is essentially similar to the gravitational interaction. As a concrete example, in~\cite{Monteiro:2020wcb}, a search for TeV-scale composite DM charged under a gauged $B-L$ symmetry was performed, leading to novel constraints with just a single $\mu$g-scale optomechanical sensor and a few days of observation time. Generically, we can parametrize the interaction potential in terms of the charge-to-mass ratio, $\lambda_{SM(DM)}$, of the SM (DM) and the new long-range coupling $g_{\phi}$. The impulse imparted on each sensor during a transit, and therefore the sensitivity of the detector, scales as $F\sim (g_{\phi}^2 \lambda_{SM} \lambda_{DM})/b^2$. Fifth-force searches generically constrain the new coupling to be weaker than $g_\phi \lesssim 10^{-21}$~\cite{Schlamminger:2007ht,Wagner:2012ui,Abbott:2021npy}. Therefore, the charge-to-mass ratio of the dark matter must be rather large, $\lambda_{DM} \gtrsim 1200 /\text{GeV}$, in order for the DM to leave visible tracks through the array~\cite{Blanco:2021yiy}. This is particularly clear when the DM is confined to a scale smaller than the typical sensor separation ($\lesssim {\rm mm-cm}$); detection prospects with less-confined, blob-like objects presents another interesting possibility \cite{Grabowska:2018lnd}.

The theory space in which the DM appears as point-like and couples through a new long-range force can be spanned by two complementary model classes. In the first, the charge confinement is realized via the formation of bound states of constituent fundamental particles. These could either belong to the SM ~\cite{Bodmer:1971we,Witten:1984rs} or to a dark sector~\cite{Bai:2018dxf,Hardy:2014mqa,Hardy:2015boa,Gresham:2017zqi,Redi:2018muu,Detmold:2014qqa,Detmold:2014kba,Gresham:2018anj,Krnjaic:2014xza,Gresham:2017cvl}. In order to be visible, these composite models must generically be strongly confining in order to evade BBN constraints. On the other hand, the dark charge can be confined in a non-fundamental particle solution to the field equations such as in non-topological solitons, i.e. Q-balls~\cite{Coleman:1985ki,Friedberg:1976me,Kusenko:1997si,Kusenko:1997zq,Kusenko:2006gv,Hong:2016ict,Hong:2020est,Holdom:1987ep,Lohiya:1994pmf,Stojkovic:2001qi,Macpherson:1994wf,Holdom:1987bu,Zhitnitsky:2002qa,Ogure:2002hv,Frieman:1988ut,Affleck:1984fy,Dine:2003ax,Krylov:2013qe}. Thick-walled Q-balls are an especially promising class of models that could be probed significantly with Windchime for charges between $10^{20}$ to $10^{30}$ and for masses above $10^{16}\;\text{GeV}$ ~\cite{Blanco:2021yiy}. In the case where the DM couples only gravitationally, the parameter space is broadly split into two regions: the sub-Planckian and super-Planckian regimes. Below the Planck scale fundamental particles can be produced gravitationally either as WIMPZILLAs around the inflationary epoch ~\cite{chung_crotty_kolb_riotto_2001, Chung:1998zb, Chung:1998ua} or by the evaporation of primordial black holes (PBHs) following an era of PBH domination~\cite{Hooper:2019gtx}. Above the Planck scale, DM can be produced in the form of Planck-scale relics of extremal gravitational singularities~\cite{Bernal:2020bjf, lehmann_johnson_profumo_schwemberger_2019, Bai:2019zcd, Aydemir:2020xfd, Abbott:2021npy,MacGibbon:1987my,Salvio:2019llz}.

\begin{figure}[t!]
\centering
$\vcenter{\hbox{\includegraphics[scale=0.4]{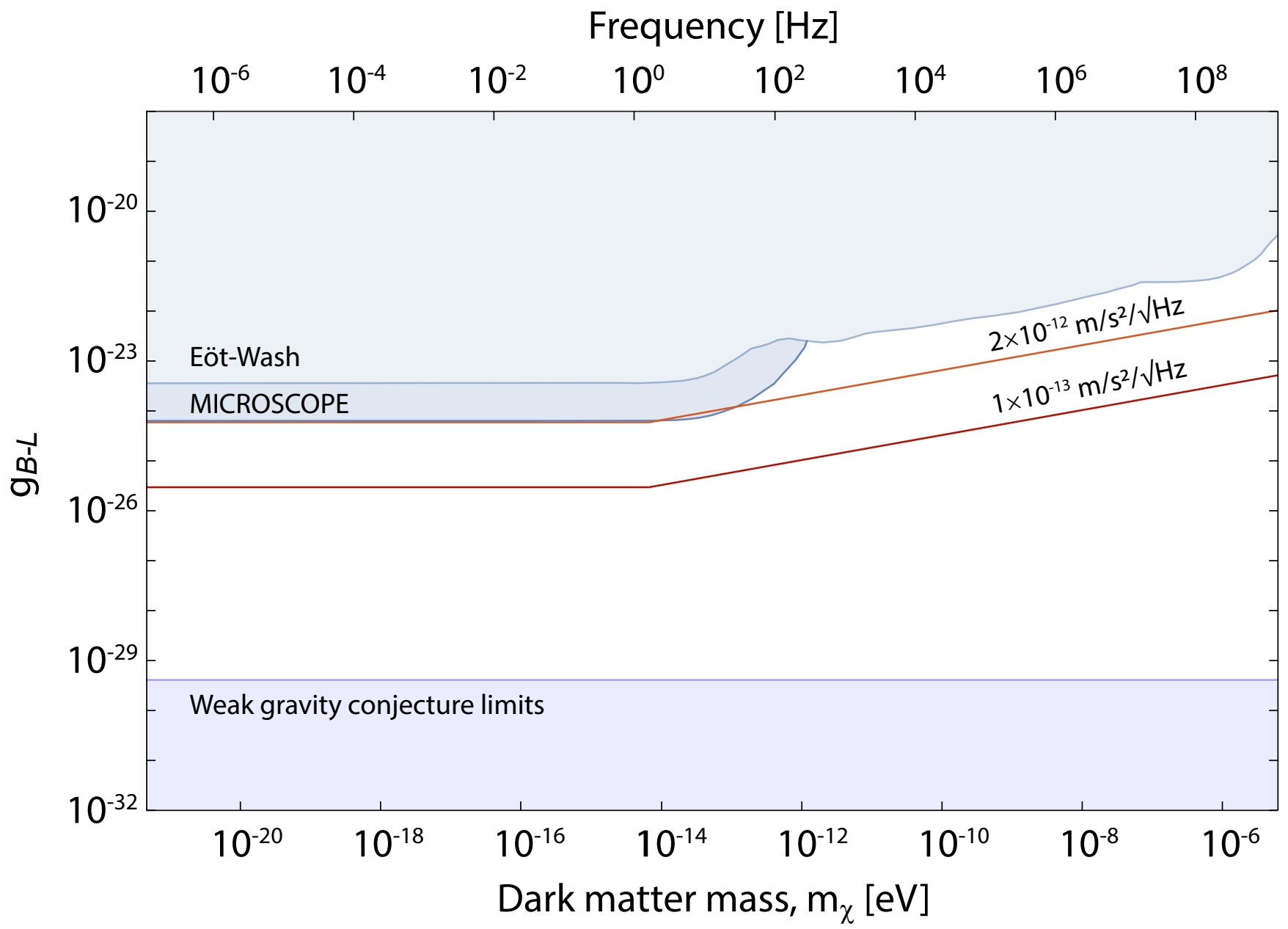}}}$ ~ ~ ~  $\vcenter{\hbox{\includegraphics[scale=0.6]{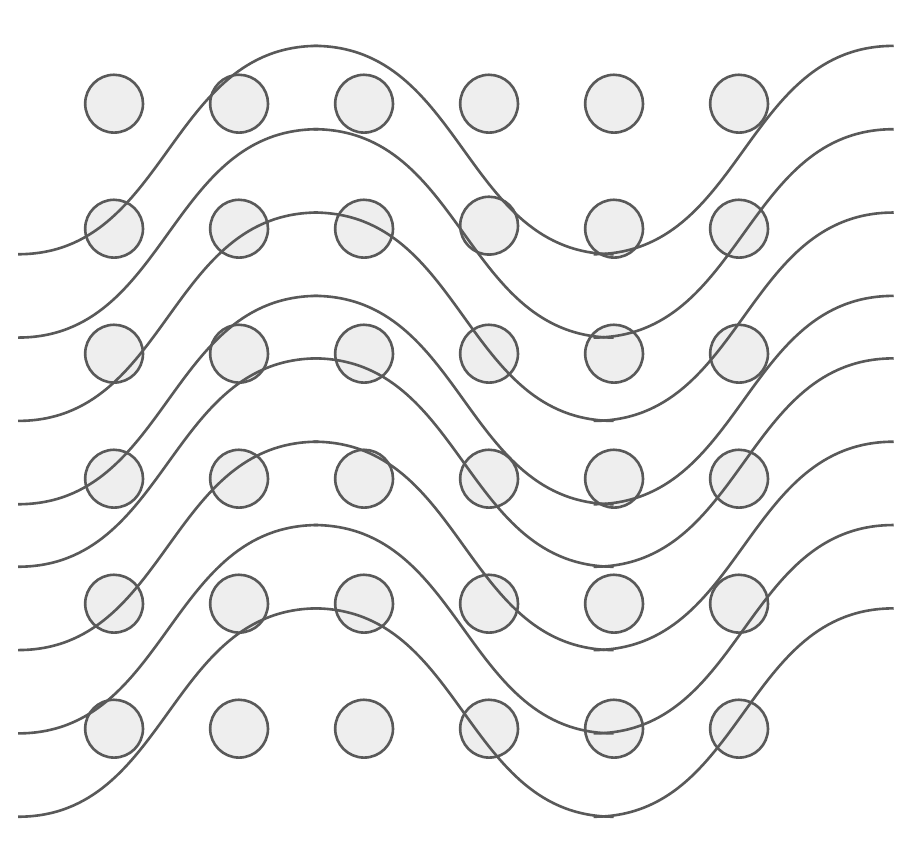}}}$
\caption{Left: Detection reach for accelerometer searches of
ultralight dark matter, taking a vector $B-L$ boson as an example. We assume one day of integration time, and the use of a pair of accelerometers with differential neutron-to-nucleon ratio $\Delta = N_1/A_1 - N_2/A_2 = 0.05$. For more details, see discussion in the text and in Refs.~\cite{graham2016dark,Carney:2019cio,carney2020mechanical}. Right: Schematic illustration of the detection signal. Here the ultralight DM field acts as a wave with coherence length larger than the linear scale of at least one sensor (a single grey blob), and possibly many sensors.}
\label{fig:ultralightprojection}
\end{figure}

Another quite different target that has been identified is  ultralight bosonic, or ``fuzzy'', dark matter \cite{SnowMassWP-CF2ULDM}. In an ultralight scenario, the DM is comprised mainly of fields with mass $m_{\chi} \lesssim 0.1~{\rm eV}$, which necessarily behaves more like a classical wave than single particle excitations due to the large number density. If this DM couples to matter, it will produce a weak, persistent, and nearly monochromatically time-varying force on each sensor. See Fig.~\ref{fig:ultralightprojection}, right panel. For example, if this ultralight field is a vector field coupling to $B-L$ charge, it will produce a force
$F(t) = g_{BL} N_{n} F_0 \sin(m_{\chi} t)$
on each sensor, where $N_{n}$ is the number of neutrons in the sensor, $g_{BL} \lesssim 10^{-24}$ is a coupling already well-constrained by fifth force experiments, and $F_0 = \sqrt{\rho_{\chi}} \approx 10^{-15}~{\rm N}$ is set by the background density of DM quanta. For $m_{\chi} \lesssim 0.1~{\rm \mu eV}$ it will act coherently over \emph{all} the sensors in the array, leading to substantial signal-to-noise enhancements. Crucially, this force scales with the mass of the sensor in a material-dependent way: it produces an acceleration sensitive to the ratio of neutrons to total nucleons of the sensor. Thus two sensors, or one sensor referenced to a fixed target, will feel a differential acceleration~\cite{graham2016dark,Carney:2019cio,carney2020mechanical}
\begin{equation}
a_{\rm signal}(t) = g_{BL} \Delta a_0 \sin(m_{\chi} t)
\end{equation}
where $\Delta = A_1/Z_1 - A_2/Z_2$ is the difference between the neutron-to-nucleon ratios of the two materials, typically a quantity of order $0.01-0.1$. Here $a_0 = F_0/m_{\rm neutron} \approx 7 \times 10^{12}~{\rm m/s^2}$. This means that we are looking for accelerations at the $10^{-12}~{\rm m/s^2}$ level. This is an ideal task for a network of quantum-limited (or enhanced) mechanical accelerometers! In Figure~\ref{fig:ultralightprojection}, we sketch the detection reach for such ultralight DM for sensors with acceleration sensitivities in this range. 
This signal can be integrated coherently over many sensors, not just a track, leading to substantial possible detection reach. 


\section{Key R\&D needs and current efforts}
\label{section-RD}

Reaching the ultimate goal of gravitational direct detection requires substantial technical development in four central directions. These are isolation from thermal environments, reduction of quantum measurement-added noise, construction and operation of a large-scale array of optomechanical sensors, and computational techniques for handling the massive data stream generated by such an array. In this section, we briefly discuss all of these issues in some detail, and some current solutions being pursued by the Windchime collaboration and others.

\subsection{Environmental isolation}

The sensors must be well isolated from thermal backgrounds in order to bring these below the weak gravitational signals of interest. With physically suspended devices, very high mechanical quality factors are of crucial importance. For an example from the current state of the art, the Matsumoto group has produced mg-scale optomechanical pendula with $Q \gtrsim 2 \times 10^6$~\cite{Catano-Lopez:2019yqb}. Moreover, the bath temperature needs to be as low as possible. Meter-scale experiments at dilution refrigeration temperatures $\sim 10~{\rm mK}$ are already demonstrated (e.g. the SuperCDMS experiment~\cite{agnese2014search}). Similar cooling methods should be adaptable to optomechanical arrays. Additionally, free-space and fiber coupled optical readout of materials and devices within dilution refrigerators at mK temperatures are also already in operation~\cite{lawrie2021free}.

Another option is to go beyond suspended systems, removing the phononic environment by levitating or even dropping the sensors. In this case, the background thermal noise comes primarily from collisions with the ambient gas molecules. Here the primary technical requirement becomes achieving a sufficiently good vacuum. This also requires cryogenic cooling, although only at the 4 K level. Pressures similar to those achieved in the LHC tunnel~\cite{grobner2001overview} appear to be sufficiently low that one could see the gravitational signal~\cite{Carney:2019pza}. More detailed study of gas damping noise spectra in the mechanical systems at these kinds of pressure, and when looking for impulsive signals, will be of crucial importance.

\subsection{Quantum noise reduction}\label{ss:QNR}

With control over technical noise sources such as stray background fields and thermal noise, mechanical sensors become limited by noise due to the quantum mechanical uncertainty of the readout system. A typical benchmark for the quantum measurement-added noise is given by the so-called Standard Quantum Limit (SQL),
\begin{equation}
\label{SQL}
\Delta p_{\rm SQL} = \sqrt{\frac{\hbar m_s}{\tau}} \approx 600~{\rm GeV/c} \times  \left( \frac{m_{s}}{1~{\rm g}} \right)^{1/2}  \left( \frac{1~{\rm \mu s}}{\tau} \right)^{1/2},
\end{equation}
where $m_s$ is the mass of the mechanical element and $\tau$ is the integration time. This is roughly set by the scale of vacuum fluctuations of the mechanical element~\cite{caves1981quantum}. Noise levels around the SQL are now routinely achieved in many systems, including the impulsive optomechanical detector of \cite{Monteiro:2020wcb}. However, SQL-level noise is orders of magnitude larger than the gravitational impulse signal~\cite{Carney:2019pza}. Thus, methods to get below the SQL level are of crucial importance to this program. Current techniques to go below the SQL are aleady in use in, e.g., advanced LIGO \cite{aasi2013enhanced}, with world records at $\sim 15 {\rm dB}$ below the SQL \cite{PhysRevLett.117.110801}. We are also developing new techniques specialized to our particular detection problems, which should have wide utility beyond the Windchime project.

A key advantage here over more typical quantum sensing problems is that the impulses are expected to be delivered relatively quickly (sub $\mu$s for typical expectations of dark matter wind velocities). Techniques for going well below the SQL can be used more readily in this case than in low frequency detection systems. Two concurrent paths have promise, and can be combined for maximum effectiveness. The first is to take advantage of squeezing of the light (or microwaves) used to measure effects of the impulse on the sensor. The role of squeezing is to trade off lower imprecision -- analogous to less shot noise -- at the price of higher backaction, in which intensity fluctuations are imprinted on the sensor and limit ultimate performance. Overall, this allows an improvement in the limiting impulse that can be measured, but only at the square root of the squeezing improvement.

The other technique is to directly monitor the velocity or momentum of the sensor, rather than infer its momentum via measurement of the position over time \cite{caves1980measurement,braginsky1990gravitational,Ghosh:2019rsc}. This enables a so-called back action evading measurement in the free mass limit -- when the detection frequency is well above the mechanical resonance of the sensor mass. In its simplest incarnation, a back action evading sensor has a velocity- or rotation rate-dependent phase shift of an optical or microwave resonance, as occurs, e.g., in a waveguide optical gyro. Improvements in phase estimation enabled by squeezing can thus be directly combined with back action evasion. To date, experimental progress in back action evasion remains nascent, with most results in the low mass ($\mu$g or smaller) regime, and thus substantial work remains to bring this to the right mass domain. Parallel to our current experimental efforts on this front, we are also actively pursuing new theoretical proposals for backaction evasion measurements tailored to impulse sensing \cite{Ghosh:2019rsc}.

\begin{figure}[b!]
\centering
\includegraphics[scale=0.5]{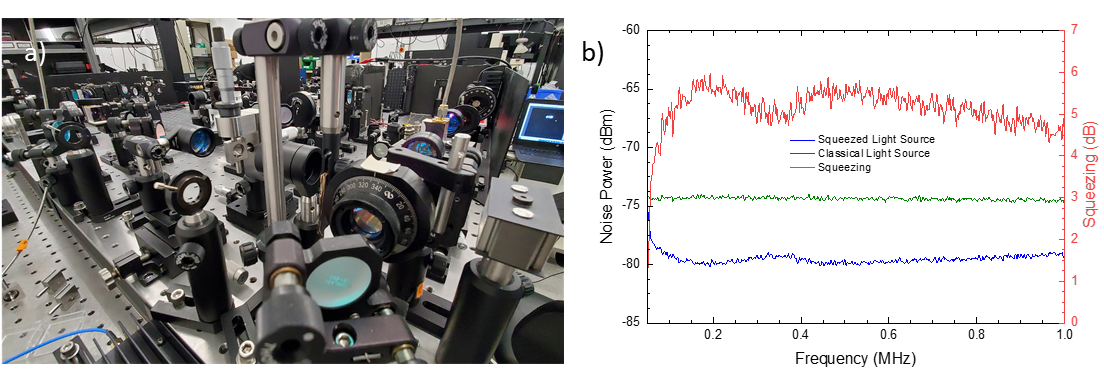}
\caption{Left: Two-mode squeezed light source apparatus, viewed from the probe arm before going through the vapor cell. Right: Noise power measurement of the two-mode squeezed light source (blue) in comparison to a shot-noise-limited classical light source (green) of comparable power, with over 5.6 dB of squeezing observed at 180 kHz.}
\label{fig:squeezing}
\end{figure}

As a first step, we are developing a squeezed optical readout technique that will be used to measure the displacement of sensors with potential for greater than 9 dB of quantum noise reduction below the shot noise limit ~\cite{glorieux2010strong}. Improved future squeezed light sources aim to reach and succeed the record in squeezing of 15 dB ~\cite{vahlbruch2016detection}. The two-mode squeezed light source used here is based on four-wave-mixing (4WM) in $^{85}$Rb, and an image of the apparatus and an example of the frequency dependent squeezing in the system is shown in Fig.~\ref{fig:squeezing}. Because the 4WM process amplifies multiple spatial modes, this approach will allow for proof-of-principle scaling of the squeezed readout field across an array of sensors \cite{lawrie2019quantum}. In this case, the sensor displacement or velocity will be read-out using a truncated nonlinear interferometer, which leverages the continuous variable entanglement in two-mode squeezed light sources to access amplitude difference or phase sum squeezing \cite{anderson2017phase}.  We have already used this detection scheme to characterize the displacement of an atomic force microscope microcantilever with quantum enhanced sensitivity \cite{pooser2020truncated,pooser2015ultrasensitive}, and we are now extending it to the domain of larger mechanical devices.

In an improved detection scenario, the sensor response may be transduced onto both channels of the two mode squeezed light source, allowing for quantum noise reduction together with a doubling of the readout signal. Because the shot noise level is defined by the total power at the detector, it is possible to improve the shot-noise-limited sensitivity by utilizing high power local oscillators with squeezed vacuum readout fields, leading to substantially reduced heating compared to schemes that rely on single mode squeezing \cite{pai2022magneto}.  This benefit is particularly critical for the ultralow temperature implementations of the sensing scheme described above. In the long term, arrays of large numbers of these sensors will be used to measure vector information, and squeezing will be used at each sensor. To accomplish this in a scalable way, waveguided sources of squeezed and entangled light will be required. We are currently developing integrated, telecom wavelength sources for this purpose, with entangled light spread over a range of frequencies that are resonant with each sensor. Including linear-optical operations in line with the resonators and sources will also allow for complex state preparation schemes in which multiparty entangled states can be shared between sensing nodes.

\subsection{Large-scale arrays}

Operating an array of many optomechanical sensors provides a unique and exciting challenge. Issues range from an appropriate choice of single-sensor architecture and readout, multiplexing, and large-scale environmental isolation. There are also substantial issues with real-time data processing, which are discussed in the next subsection.

As a first step, we are constructing a small array of milligram-scale accelerometers named \emph{Protochime}. This prototype is an array of 15 commercial MEMS (microelectronic mechanical systems) accelerometers, set up to take data continuously and in parallel for 16 channels with sampling rates in the order of low MHz. The sensors are mechanically mounted on one or more solid material blocks with sizes in the cm-m range. This array is able to measure correlated vibrations produced by external driving forces along the sensitive axis if they are above a certain noise level determined by the accelerometer noise level, the electronics and the bit-precision in the digitization process of the data. Successful measurements with this prototype include a proof-of-principle constraint (not competitive with pre-existing fifth force bounds) on ultralight DM models, following the general methodology proposed in~\cite{graham2016dark,Carney:2019cio}. Protochime serves as the first experimental realization capable of producing real data to test the novel approach of Windchime in DM detection. It is useful to explore current hardware and software scope and limitations. 

Laser shot-noise limited optomechanical accelerometers have been demonstrated by coupling optical free-space readout of suspended proof-masses \cite{zhou2021, guzman2014}. However, Windchime requires at least $10^6$ such accelerometers. A frequency multiplexed and optically coherent readout scheme is thus necessary. One potential solution is to interface the extremely sensitive MEMS \cite{hutchison2012z} with photonic integrated circuits (PIC) and take advantage of the frequency tuning \cite{tian2020}, multiplexing, combs \cite{liu2020} and coherent readout technologies demonstrated on PIC platforms. The next generation array that we are working on will use $\sim 10^2$ dedicated accelerometers based on integrated Si$_3$N$_4$ photonic circuits, achieving a noise density in the order of ${\rm \mu g}/\sqrt{\rm Hz}$ at room temperature and pressure. These sensors are based around an optical microring resonator, and would be read out using a fibre-coupled inteferometric setup \cite{krause2012}. The optical resonance frequency is adjustable on the order of GHz by integrating piezoelectric actuators \cite{liu2020}, allowing for frequency domain multiplexing. A single frequency laser can be electro-optically modulated for producing a frequency comb \cite{zhang2019}, where each comb line can be aligned with each sensor with low cross-talk. Such a setup will serve as a test-bed for the technological developments needed to detect dark matter with an optomechanical array.

\subsection{Computational techniques}

Windchime presents interesting R\&D opportunities as we develop the computational techniques needed for large arrays of quantum-enabled mechanical sensors. The array produces streaming spatiotemporal data, where each sensor is producing its own time series data. The novelty of Windchime is in using a large number of sensors to take advantage of incoherent noise.  Beyond the challenges of having many sensors, the data analysis presents its own challenges. However, there is a substantial body of previous work on related problems; as a benchmark, the CMS silicon tracker has 10 million readout channels being sampled at a rate of 40 MHz, then read out at up to 100 kHz~\cite{Foudas:2004axy}.

There are two different scientific objectives that provide competing requirements for how signals could be detected in such an array. Heavy dark matter will produce a fast transient signal in the array with signals localized in a few sensors, which requires a faster digitization frequency thus high date rate per sensor. However, ultralight dark matter is a coherent effect across the entire array, where information from all sensors is needed to search for a signal, thereby scaling the high sensor rate by a large number of sensors. Therefore, different techniques are required for both of these extremes.

For heavy dark matter, we are performing track finding in an optomechanical sensor array to identify sensors nearby the track.  Track finding is a mature field in particle physics, as this is a core part of the analysis of most collider physics detectors. The main computational challenges related to this task are areas of active research in computational science. First, how does one perform triggering, which is also known as object detection, to only record interesting data? This requires having streaming algorithms to identify where they may be a track. Second, how does one perform the analysis or optimization to determine where a track might be? An interesting possibility is to use signal processing techniques to preprocess this data. Third, what opportunities are there to quantify uncertainties on the track parameters, such as is done for sky localization in multimessenger astronomy?

Because gravity is a long range interaction, even sensors that are not directly impacted by the track can contribute to a signal. This means that parameter inference could be more sophisticated than basic track finding, for example using Bayesian methods to infer the posterior distribution of the track from data. This has been successfully trialed using a simulated array of \(8^3\) sensors and the ultranest package~\cite{2021JOSS....6.3001B}. Such methods, however, would be highly computationally intensive, necessitating the use of a trigger as mentioned above. In addition, triggering might be needed for data storage reasons. For example, an array of \(10^6\) sensors sampled at \(10 \, \mathrm{MHz}\) at 12 bits would result in a data rate of about \(15 \, \mathrm{TB/s}\). Only storing triggered data will be crucial to reduce the data storage requirements. 

For ultralight dark matter, the noise level can be suppressed by up to $\sqrt{N_{det}}$, as discussed above. The main challenge is designing a data acquisition system that can probe coherent effects across the entire array.  The ultralight dark matter signal has a finite coherence time, inversely proportional to its bandwidth, $T_{coh}\sim10^6/\omega_{DM}$. Therefore, for the experiment time scale that is shorter than the dark matter coherence time (e.g., $T_{coh}>$1yr for $m_{DM}<10^{-13}$eV), the signal behaves as a monochromatic wave with a single frequency across all sensors.  In that case, a time-domain Wiener optimal filter can be used (even at the FPGA level). For larger DM mass, thus smaller coherence time, the DM produces an incoherent power signal. This may require offline processing of the time series, potentially with a method similar to a Dicke radiometer, whose analysis framework was presented in~\cite{Chaudhuri:2018rqn}, while for smaller timescales analysis procedures similar to the one presented in ~\cite{Salemi:2021} can be used.

In all cases above, an exciting opportunity is performing machine-learning-enabled experimental design.  As we are operating a cubic lattice, simulations and forward models of the processes involved lend themselves to optimization techniques that operate on large matrices.  Existing routines for analyzing data on a grid, such as convolutional neural networks, mean that there is an exciting opportunity for being able to perform global detector optimization in a way that is not as tractable in other particle physics experiments.

\section{Outlook}

The direct detection of dark matter through gravity alone was long thought to be impossible. Thanks to advances made in the control of optomechanical systems, we have identified a possible avenue to achieve this goal in the foreseeable future. While certainly a long-term endeavour, the Windchime project is charting a path towards achieving sensitivity to well-motivated Planck-mass dark matter through gravitational interaction alone. On the way to such sensitivity to this well-motivated parameter space, Windchime is also probing ultralight DM candidates, and possibly a number of other targets of interest. A number of applications of this technology to other areas---including acceleration sensing, impulse metrology, and multiplexing of large quantum sensing arrays---can be expected in the process. Taken together, the Windchime project constitutes a unique and well-motivated approach to enhance our sensitivity to particle dark matter.

\newpage

\bibliography{snowmass-windchime}

\begin{thebibliography}{92}%
\makeatletter
\providecommand \@ifxundefined [1]{%
 \@ifx{#1\undefined}
}%
\providecommand \@ifnum [1]{%
 \ifnum #1\expandafter \@firstoftwo
 \else \expandafter \@secondoftwo
 \fi
}%
\providecommand \@ifx [1]{%
 \ifx #1\expandafter \@firstoftwo
 \else \expandafter \@secondoftwo
 \fi
}%
\providecommand \natexlab [1]{#1}%
\providecommand \enquote  [1]{``#1''}%
\providecommand \bibnamefont  [1]{#1}%
\providecommand \bibfnamefont [1]{#1}%
\providecommand \citenamefont [1]{#1}%
\providecommand \href@noop [0]{\@secondoftwo}%
\providecommand \href [0]{\begingroup \@sanitize@url \@href}%
\providecommand \@href[1]{\@@startlink{#1}\@@href}%
\providecommand \@@href[1]{\endgroup#1\@@endlink}%
\providecommand \@sanitize@url [0]{\catcode `\\12\catcode `\$12\catcode
  `\&12\catcode `\#12\catcode `\^12\catcode `\_12\catcode `\%12\relax}%
\providecommand \@@startlink[1]{}%
\providecommand \@@endlink[0]{}%
\providecommand \url  [0]{\begingroup\@sanitize@url \@url }%
\providecommand \@url [1]{\endgroup\@href {#1}{\urlprefix }}%
\providecommand \urlprefix  [0]{URL }%
\providecommand \Eprint [0]{\href }%
\providecommand \doibase [0]{http://dx.doi.org/}%
\providecommand \selectlanguage [0]{\@gobble}%
\providecommand \bibinfo  [0]{\@secondoftwo}%
\providecommand \bibfield  [0]{\@secondoftwo}%
\providecommand \translation [1]{[#1]}%
\providecommand \BibitemOpen [0]{}%
\providecommand \bibitemStop [0]{}%
\providecommand \bibitemNoStop [0]{.\EOS\space}%
\providecommand \EOS [0]{\spacefactor3000\relax}%
\providecommand \BibitemShut  [1]{\csname bibitem#1\endcsname}%
\let\auto@bib@innerbib\@empty
\bibitem [{\citenamefont {Hall}\ \emph {et~al.}(2018)\citenamefont {Hall},
  \citenamefont {Adhikari}, \citenamefont {Frolov}, \citenamefont {M\"uller},
  \citenamefont {Pospelov},\ and\ \citenamefont {Adhikari}}]{Hall:2016usm}%
  \BibitemOpen
  \bibfield  {author} {\bibinfo {author} {\bibfnamefont {E.~D.}\ \bibnamefont
  {Hall}}, \bibinfo {author} {\bibfnamefont {R.~X.}\ \bibnamefont {Adhikari}},
  \bibinfo {author} {\bibfnamefont {V.~V.}\ \bibnamefont {Frolov}}, \bibinfo
  {author} {\bibfnamefont {H.}~\bibnamefont {M\"uller}}, \bibinfo {author}
  {\bibfnamefont {M.}~\bibnamefont {Pospelov}}, \ and\ \bibinfo {author}
  {\bibfnamefont {R.~X.}\ \bibnamefont {Adhikari}},\ }\href {\doibase
  10.1103/PhysRevD.98.083019} {\bibfield  {journal} {\bibinfo  {journal} {Phys.
  Rev. D}\ }\textbf {\bibinfo {volume} {98}},\ \bibinfo {pages} {083019}
  (\bibinfo {year} {2018})},\ \Eprint {http://arxiv.org/abs/1605.01103}
  {arXiv:1605.01103 [gr-qc]} \BibitemShut {NoStop}%
\bibitem [{\citenamefont {Kawasaki}(2019)}]{Kawasaki:2018xak}%
  \BibitemOpen
  \bibfield  {author} {\bibinfo {author} {\bibfnamefont {A.}~\bibnamefont
  {Kawasaki}},\ }\href {\doibase 10.1103/PhysRevD.99.023005} {\bibfield
  {journal} {\bibinfo  {journal} {Phys. Rev. D}\ }\textbf {\bibinfo {volume}
  {99}},\ \bibinfo {pages} {023005} (\bibinfo {year} {2019})},\ \Eprint
  {http://arxiv.org/abs/1809.00968} {arXiv:1809.00968 [physics.ins-det]}
  \BibitemShut {NoStop}%
\bibitem [{\citenamefont {Carney}\ \emph {et~al.}(2020)\citenamefont {Carney},
  \citenamefont {Ghosh}, \citenamefont {Krnjaic},\ and\ \citenamefont
  {Taylor}}]{Carney:2019pza}%
  \BibitemOpen
  \bibfield  {author} {\bibinfo {author} {\bibfnamefont {D.}~\bibnamefont
  {Carney}}, \bibinfo {author} {\bibfnamefont {S.}~\bibnamefont {Ghosh}},
  \bibinfo {author} {\bibfnamefont {G.}~\bibnamefont {Krnjaic}}, \ and\
  \bibinfo {author} {\bibfnamefont {J.~M.}\ \bibnamefont {Taylor}},\ }\href
  {\doibase 10.1103/PhysRevD.102.072003} {\bibfield  {journal} {\bibinfo
  {journal} {Phys. Rev. D}\ }\textbf {\bibinfo {volume} {102}},\ \bibinfo
  {pages} {072003} (\bibinfo {year} {2020})},\ \Eprint
  {http://arxiv.org/abs/1903.00492} {arXiv:1903.00492 [hep-ph]} \BibitemShut
  {NoStop}%
\bibitem [{\citenamefont {Aspelmeyer}\ \emph {et~al.}(2014)\citenamefont
  {Aspelmeyer}, \citenamefont {Kippenberg},\ and\ \citenamefont
  {Marquardt}}]{aspelmeyer2014cavity}%
  \BibitemOpen
  \bibfield  {author} {\bibinfo {author} {\bibfnamefont {M.}~\bibnamefont
  {Aspelmeyer}}, \bibinfo {author} {\bibfnamefont {T.~J.}\ \bibnamefont
  {Kippenberg}}, \ and\ \bibinfo {author} {\bibfnamefont {F.}~\bibnamefont
  {Marquardt}},\ }\href@noop {} {\bibfield  {journal} {\bibinfo  {journal}
  {Reviews of Modern Physics}\ }\textbf {\bibinfo {volume} {86}},\ \bibinfo
  {pages} {1391} (\bibinfo {year} {2014})}\BibitemShut {NoStop}%
\bibitem [{\citenamefont {Blencowe}(2004)}]{blencowe2004quantum}%
  \BibitemOpen
  \bibfield  {author} {\bibinfo {author} {\bibfnamefont {M.}~\bibnamefont
  {Blencowe}},\ }\href@noop {} {\bibfield  {journal} {\bibinfo  {journal}
  {Physics Reports}\ }\textbf {\bibinfo {volume} {395}},\ \bibinfo {pages}
  {159} (\bibinfo {year} {2004})}\BibitemShut {NoStop}%
\bibitem [{\citenamefont {{Snowmass 2021 Instrumentation Frontier: Quantum
  Sensing for HEP Science --- Interferometers, Mechanics, Clocks, and
  Traps}}()}]{SnowMassWP-IFquantum}%
  \BibitemOpen
  \bibfield  {author} {\bibinfo {author} {\bibnamefont {{Snowmass 2021
  Instrumentation Frontier: Quantum Sensing for HEP Science ---
  Interferometers, Mechanics, Clocks, and Traps}}},\ }\href@noop {} {\ }\Eprint
  {http://arxiv.org/abs/2203.xxxxx} {arXiv:2203.xxxxx} \BibitemShut {NoStop}%
\bibitem [{\citenamefont {Carney}\ \emph {et~al.}()\citenamefont {Carney},
  \citenamefont {Krnjaic}, \citenamefont {Moore}, \citenamefont {Regal} \emph
  {et~al.}}]{carney2020mechanical}%
  \BibitemOpen
  \bibfield  {author} {\bibinfo {author} {\bibfnamefont {D.}~\bibnamefont
  {Carney}}, \bibinfo {author} {\bibfnamefont {G.}~\bibnamefont {Krnjaic}},
  \bibinfo {author} {\bibfnamefont {D.~C.}\ \bibnamefont {Moore}}, \bibinfo
  {author} {\bibfnamefont {C.~A.}\ \bibnamefont {Regal}},  \emph {et~al.},\
  }\href@noop {} {\ }\Eprint {http://arxiv.org/abs/2008.06074}
  {arXiv:2008.06074 [physics.ins-det]} \BibitemShut {NoStop}%
\bibitem [{\citenamefont {Graham}\ \emph {et~al.}(2016)\citenamefont {Graham},
  \citenamefont {Kaplan}, \citenamefont {Mardon}, \citenamefont {Rajendran},\
  and\ \citenamefont {Terrano}}]{graham2016dark}%
  \BibitemOpen
  \bibfield  {author} {\bibinfo {author} {\bibfnamefont {P.~W.}\ \bibnamefont
  {Graham}}, \bibinfo {author} {\bibfnamefont {D.~E.}\ \bibnamefont {Kaplan}},
  \bibinfo {author} {\bibfnamefont {J.}~\bibnamefont {Mardon}}, \bibinfo
  {author} {\bibfnamefont {S.}~\bibnamefont {Rajendran}}, \ and\ \bibinfo
  {author} {\bibfnamefont {W.~A.}\ \bibnamefont {Terrano}},\ }\href@noop {}
  {\bibfield  {journal} {\bibinfo  {journal} {Physical Review D}\ }\textbf
  {\bibinfo {volume} {93}},\ \bibinfo {pages} {075029} (\bibinfo {year}
  {2016})}\BibitemShut {NoStop}%
\bibitem [{\citenamefont {Carney}\ \emph {et~al.}(2019)\citenamefont {Carney},
  \citenamefont {Hook}, \citenamefont {Liu}, \citenamefont {Taylor},\ and\
  \citenamefont {Zhao}}]{Carney:2019cio}%
  \BibitemOpen
  \bibfield  {author} {\bibinfo {author} {\bibfnamefont {D.}~\bibnamefont
  {Carney}}, \bibinfo {author} {\bibfnamefont {A.}~\bibnamefont {Hook}},
  \bibinfo {author} {\bibfnamefont {Z.}~\bibnamefont {Liu}}, \bibinfo {author}
  {\bibfnamefont {J.~M.}\ \bibnamefont {Taylor}}, \ and\ \bibinfo {author}
  {\bibfnamefont {Y.}~\bibnamefont {Zhao}},\ }\href@noop {} {\  (\bibinfo
  {year} {2019})},\ \Eprint {http://arxiv.org/abs/1908.04797} {arXiv:1908.04797
  [hep-ph]} \BibitemShut {NoStop}%
\bibitem [{\citenamefont {Manley}\ \emph {et~al.}(2020)\citenamefont {Manley},
  \citenamefont {Chowdhury}, \citenamefont {Grin}, \citenamefont {Singh},\ and\
  \citenamefont {Wilson}}]{Manley:2020mjq}%
  \BibitemOpen
  \bibfield  {author} {\bibinfo {author} {\bibfnamefont {J.}~\bibnamefont
  {Manley}}, \bibinfo {author} {\bibfnamefont {M.~D.}\ \bibnamefont
  {Chowdhury}}, \bibinfo {author} {\bibfnamefont {D.}~\bibnamefont {Grin}},
  \bibinfo {author} {\bibfnamefont {S.}~\bibnamefont {Singh}}, \ and\ \bibinfo
  {author} {\bibfnamefont {D.~J.}\ \bibnamefont {Wilson}},\ }\href@noop {} {\
  (\bibinfo {year} {2020})},\ \Eprint {http://arxiv.org/abs/2007.04899}
  {arXiv:2007.04899 [quant-ph]} \BibitemShut {NoStop}%
\bibitem [{\citenamefont {{Snowmass 2021 CF2 Whitepaper: New Horizons: Scalar
  and Vector Ultralight Dark Matter}}()}]{SnowMassWP-CF2ULDM}%
  \BibitemOpen
  \bibfield  {author} {\bibinfo {author} {\bibnamefont {{Snowmass 2021 CF2
  Whitepaper: New Horizons: Scalar and Vector Ultralight Dark Matter}}},\
  }\href@noop {} {\ }\Eprint {http://arxiv.org/abs/2203.xxxxx}
  {arXiv:2203.xxxxx} \BibitemShut {NoStop}%
\bibitem [{\citenamefont {Afek}\ \emph {et~al.}(2022)\citenamefont {Afek},
  \citenamefont {Carney},\ and\ \citenamefont {Moore}}]{Afek:2021vjy}%
  \BibitemOpen
  \bibfield  {author} {\bibinfo {author} {\bibfnamefont {G.}~\bibnamefont
  {Afek}}, \bibinfo {author} {\bibfnamefont {D.}~\bibnamefont {Carney}}, \ and\
  \bibinfo {author} {\bibfnamefont {D.~C.}\ \bibnamefont {Moore}},\ }\href
  {\doibase 10.1103/PhysRevLett.128.101301} {\bibfield  {journal} {\bibinfo
  {journal} {Phys. Rev. Lett.}\ }\textbf {\bibinfo {volume} {128}},\ \bibinfo
  {pages} {101301} (\bibinfo {year} {2022})},\ \Eprint
  {http://arxiv.org/abs/2111.03597} {arXiv:2111.03597 [physics.ins-det]}
  \BibitemShut {NoStop}%
\bibitem [{\citenamefont {Monteiro}\ \emph {et~al.}(2020)\citenamefont
  {Monteiro}, \citenamefont {Afek}, \citenamefont {Carney}, \citenamefont
  {Krnjaic}, \citenamefont {Wang},\ and\ \citenamefont
  {Moore}}]{Monteiro:2020wcb}%
  \BibitemOpen
  \bibfield  {author} {\bibinfo {author} {\bibfnamefont {F.}~\bibnamefont
  {Monteiro}}, \bibinfo {author} {\bibfnamefont {G.}~\bibnamefont {Afek}},
  \bibinfo {author} {\bibfnamefont {D.}~\bibnamefont {Carney}}, \bibinfo
  {author} {\bibfnamefont {G.}~\bibnamefont {Krnjaic}}, \bibinfo {author}
  {\bibfnamefont {J.}~\bibnamefont {Wang}}, \ and\ \bibinfo {author}
  {\bibfnamefont {D.~C.}\ \bibnamefont {Moore}},\ }\href@noop {} {\  (\bibinfo
  {year} {2020})},\ \Eprint {http://arxiv.org/abs/2007.12067} {arXiv:2007.12067
  [hep-ex]} \BibitemShut {NoStop}%
\bibitem [{\citenamefont {{Snowmass 2021 Cosmic Frontier White Paper:
  Ultraheavy dark matter}}()}]{SnowMassWP-CF1UHDM}%
  \BibitemOpen
  \bibfield  {author} {\bibinfo {author} {\bibnamefont {{Snowmass 2021 Cosmic
  Frontier White Paper: Ultraheavy dark matter}}},\ }\href@noop {} {\ }\Eprint
  {http://arxiv.org/abs/2203.xxxxx} {arXiv:2203.xxxxx} \BibitemShut {NoStop}%
\bibitem [{\citenamefont {Abramovici}\ \emph {et~al.}(1992)\citenamefont
  {Abramovici}, \citenamefont {Althouse}, \citenamefont {Drever}, \citenamefont
  {G{\"u}rsel}, \citenamefont {Kawamura}, \citenamefont {Raab}, \citenamefont
  {Shoemaker}, \citenamefont {Sievers}, \citenamefont {Spero}, \citenamefont
  {Thorne} \emph {et~al.}}]{abramovici1992ligo}%
  \BibitemOpen
  \bibfield  {author} {\bibinfo {author} {\bibfnamefont {A.}~\bibnamefont
  {Abramovici}}, \bibinfo {author} {\bibfnamefont {W.~E.}\ \bibnamefont
  {Althouse}}, \bibinfo {author} {\bibfnamefont {R.~W.}\ \bibnamefont
  {Drever}}, \bibinfo {author} {\bibfnamefont {Y.}~\bibnamefont {G{\"u}rsel}},
  \bibinfo {author} {\bibfnamefont {S.}~\bibnamefont {Kawamura}}, \bibinfo
  {author} {\bibfnamefont {F.~J.}\ \bibnamefont {Raab}}, \bibinfo {author}
  {\bibfnamefont {D.}~\bibnamefont {Shoemaker}}, \bibinfo {author}
  {\bibfnamefont {L.}~\bibnamefont {Sievers}}, \bibinfo {author} {\bibfnamefont
  {R.~E.}\ \bibnamefont {Spero}}, \bibinfo {author} {\bibfnamefont {K.~S.}\
  \bibnamefont {Thorne}},  \emph {et~al.},\ }\href@noop {} {\bibfield
  {journal} {\bibinfo  {journal} {Science}\ }\textbf {\bibinfo {volume}
  {256}},\ \bibinfo {pages} {325} (\bibinfo {year} {1992})}\BibitemShut
  {NoStop}%
\bibitem [{\citenamefont {Cataño-Lopez}\ \emph {et~al.}(2020)\citenamefont
  {Cataño-Lopez}, \citenamefont {Santiago-Condori}, \citenamefont {Edamatsu},\
  and\ \citenamefont {Matsumoto}}]{Catano-Lopez:2019yqb}%
  \BibitemOpen
  \bibfield  {author} {\bibinfo {author} {\bibfnamefont {S.~B.}\ \bibnamefont
  {Cataño-Lopez}}, \bibinfo {author} {\bibfnamefont {J.~G.}\ \bibnamefont
  {Santiago-Condori}}, \bibinfo {author} {\bibfnamefont {K.}~\bibnamefont
  {Edamatsu}}, \ and\ \bibinfo {author} {\bibfnamefont {N.}~\bibnamefont
  {Matsumoto}},\ }\href {\doibase 10.1103/PhysRevLett.124.221102} {\bibfield
  {journal} {\bibinfo  {journal} {Phys. Rev. Lett.}\ }\textbf {\bibinfo
  {volume} {124}},\ \bibinfo {pages} {221102} (\bibinfo {year} {2020})},\
  \Eprint {http://arxiv.org/abs/1912.12567} {arXiv:1912.12567 [quant-ph]}
  \BibitemShut {NoStop}%
\bibitem [{\citenamefont {Deli{\'c}}\ \emph {et~al.}(2020)\citenamefont
  {Deli{\'c}}, \citenamefont {Reisenbauer}, \citenamefont {Dare}, \citenamefont
  {Grass}, \citenamefont {Vuleti{\'c}}, \citenamefont {Kiesel},\ and\
  \citenamefont {Aspelmeyer}}]{delic2020cooling}%
  \BibitemOpen
  \bibfield  {author} {\bibinfo {author} {\bibfnamefont {U.}~\bibnamefont
  {Deli{\'c}}}, \bibinfo {author} {\bibfnamefont {M.}~\bibnamefont
  {Reisenbauer}}, \bibinfo {author} {\bibfnamefont {K.}~\bibnamefont {Dare}},
  \bibinfo {author} {\bibfnamefont {D.}~\bibnamefont {Grass}}, \bibinfo
  {author} {\bibfnamefont {V.}~\bibnamefont {Vuleti{\'c}}}, \bibinfo {author}
  {\bibfnamefont {N.}~\bibnamefont {Kiesel}}, \ and\ \bibinfo {author}
  {\bibfnamefont {M.}~\bibnamefont {Aspelmeyer}},\ }\href@noop {} {\bibfield
  {journal} {\bibinfo  {journal} {Science}\ }\textbf {\bibinfo {volume}
  {367}},\ \bibinfo {pages} {892} (\bibinfo {year} {2020})}\BibitemShut
  {NoStop}%
\bibitem [{\citenamefont {Moody}\ \emph {et~al.}(2002)\citenamefont {Moody},
  \citenamefont {Paik},\ and\ \citenamefont {Canavan}}]{moody2002three}%
  \BibitemOpen
  \bibfield  {author} {\bibinfo {author} {\bibfnamefont {M.~V.}\ \bibnamefont
  {Moody}}, \bibinfo {author} {\bibfnamefont {H.~J.}\ \bibnamefont {Paik}}, \
  and\ \bibinfo {author} {\bibfnamefont {E.~R.}\ \bibnamefont {Canavan}},\
  }\href@noop {} {\bibfield  {journal} {\bibinfo  {journal} {Review of
  scientific instruments}\ }\textbf {\bibinfo {volume} {73}},\ \bibinfo {pages}
  {3957} (\bibinfo {year} {2002})}\BibitemShut {NoStop}%
\bibitem [{\citenamefont {Romero-Isart}\ \emph {et~al.}(2012)\citenamefont
  {Romero-Isart}, \citenamefont {Clemente}, \citenamefont {Navau},
  \citenamefont {Sanchez},\ and\ \citenamefont {Cirac}}]{romero2012quantum}%
  \BibitemOpen
  \bibfield  {author} {\bibinfo {author} {\bibfnamefont {O.}~\bibnamefont
  {Romero-Isart}}, \bibinfo {author} {\bibfnamefont {L.}~\bibnamefont
  {Clemente}}, \bibinfo {author} {\bibfnamefont {C.}~\bibnamefont {Navau}},
  \bibinfo {author} {\bibfnamefont {A.}~\bibnamefont {Sanchez}}, \ and\
  \bibinfo {author} {\bibfnamefont {J.}~\bibnamefont {Cirac}},\ }\href@noop {}
  {\bibfield  {journal} {\bibinfo  {journal} {Physical review letters}\
  }\textbf {\bibinfo {volume} {109}},\ \bibinfo {pages} {147205} (\bibinfo
  {year} {2012})}\BibitemShut {NoStop}%
\bibitem [{\citenamefont {Bramante}\ \emph {et~al.}(2018)\citenamefont
  {Bramante}, \citenamefont {Broerman}, \citenamefont {Lang},\ and\
  \citenamefont {Raj}}]{Bramante:2018qbc}%
  \BibitemOpen
  \bibfield  {author} {\bibinfo {author} {\bibfnamefont {J.}~\bibnamefont
  {Bramante}}, \bibinfo {author} {\bibfnamefont {B.}~\bibnamefont {Broerman}},
  \bibinfo {author} {\bibfnamefont {R.~F.}\ \bibnamefont {Lang}}, \ and\
  \bibinfo {author} {\bibfnamefont {N.}~\bibnamefont {Raj}},\ }\href {\doibase
  10.1103/PhysRevD.98.083516} {\bibfield  {journal} {\bibinfo  {journal} {Phys.
  Rev. D}\ }\textbf {\bibinfo {volume} {98}},\ \bibinfo {pages} {083516}
  (\bibinfo {year} {2018})},\ \Eprint {http://arxiv.org/abs/1803.08044}
  {arXiv:1803.08044 [hep-ph]} \BibitemShut {NoStop}%
\bibitem [{\citenamefont {Schlamminger}\ \emph {et~al.}(2008)\citenamefont
  {Schlamminger}, \citenamefont {Choi}, \citenamefont {Wagner}, \citenamefont
  {Gundlach},\ and\ \citenamefont {Adelberger}}]{Schlamminger:2007ht}%
  \BibitemOpen
  \bibfield  {author} {\bibinfo {author} {\bibfnamefont {S.}~\bibnamefont
  {Schlamminger}}, \bibinfo {author} {\bibfnamefont {K.~Y.}\ \bibnamefont
  {Choi}}, \bibinfo {author} {\bibfnamefont {T.~A.}\ \bibnamefont {Wagner}},
  \bibinfo {author} {\bibfnamefont {J.~H.}\ \bibnamefont {Gundlach}}, \ and\
  \bibinfo {author} {\bibfnamefont {E.~G.}\ \bibnamefont {Adelberger}},\ }\href
  {\doibase 10.1103/PhysRevLett.100.041101} {\bibfield  {journal} {\bibinfo
  {journal} {Phys. Rev. Lett.}\ }\textbf {\bibinfo {volume} {100}},\ \bibinfo
  {pages} {041101} (\bibinfo {year} {2008})},\ \Eprint
  {http://arxiv.org/abs/0712.0607} {arXiv:0712.0607 [gr-qc]} \BibitemShut
  {NoStop}%
\bibitem [{\citenamefont {Wagner}\ \emph {et~al.}(2012)\citenamefont {Wagner},
  \citenamefont {Schlamminger}, \citenamefont {Gundlach},\ and\ \citenamefont
  {Adelberger}}]{Wagner:2012ui}%
  \BibitemOpen
  \bibfield  {author} {\bibinfo {author} {\bibfnamefont {T.~A.}\ \bibnamefont
  {Wagner}}, \bibinfo {author} {\bibfnamefont {S.}~\bibnamefont
  {Schlamminger}}, \bibinfo {author} {\bibfnamefont {J.~H.}\ \bibnamefont
  {Gundlach}}, \ and\ \bibinfo {author} {\bibfnamefont {E.~G.}\ \bibnamefont
  {Adelberger}},\ }\href {\doibase 10.1088/0264-9381/29/18/184002} {\bibfield
  {journal} {\bibinfo  {journal} {Class. Quant. Grav.}\ }\textbf {\bibinfo
  {volume} {29}},\ \bibinfo {pages} {184002} (\bibinfo {year} {2012})},\
  \Eprint {http://arxiv.org/abs/1207.2442} {arXiv:1207.2442 [gr-qc]}
  \BibitemShut {NoStop}%
\bibitem [{\citenamefont {Abbott}\ \emph {et~al.}(2021)\citenamefont {Abbott}
  \emph {et~al.}}]{Abbott:2021npy}%
  \BibitemOpen
  \bibfield  {author} {\bibinfo {author} {\bibfnamefont {R.}~\bibnamefont
  {Abbott}} \emph {et~al.} (\bibinfo {collaboration} {LIGO Scientific, Virgo,
  KAGRA}),\ }\href@noop {} {\  (\bibinfo {year} {2021})},\ \Eprint
  {http://arxiv.org/abs/2105.13085} {arXiv:2105.13085 [astro-ph.CO]}
  \BibitemShut {NoStop}%
\bibitem [{\citenamefont {Blanco}\ \emph {et~al.}(2021)\citenamefont {Blanco},
  \citenamefont {Elshimy}, \citenamefont {Lang},\ and\ \citenamefont
  {Orlando}}]{Blanco:2021yiy}%
  \BibitemOpen
  \bibfield  {author} {\bibinfo {author} {\bibfnamefont {C.}~\bibnamefont
  {Blanco}}, \bibinfo {author} {\bibfnamefont {B.}~\bibnamefont {Elshimy}},
  \bibinfo {author} {\bibfnamefont {R.~F.}\ \bibnamefont {Lang}}, \ and\
  \bibinfo {author} {\bibfnamefont {R.}~\bibnamefont {Orlando}},\ }\href@noop
  {} {\  (\bibinfo {year} {2021})},\ \Eprint {http://arxiv.org/abs/2112.14784}
  {arXiv:2112.14784 [hep-ph]} \BibitemShut {NoStop}%
\bibitem [{\citenamefont {Grabowska}\ \emph {et~al.}(2018)\citenamefont
  {Grabowska}, \citenamefont {Melia},\ and\ \citenamefont
  {Rajendran}}]{Grabowska:2018lnd}%
  \BibitemOpen
  \bibfield  {author} {\bibinfo {author} {\bibfnamefont {D.~M.}\ \bibnamefont
  {Grabowska}}, \bibinfo {author} {\bibfnamefont {T.}~\bibnamefont {Melia}}, \
  and\ \bibinfo {author} {\bibfnamefont {S.}~\bibnamefont {Rajendran}},\ }\href
  {\doibase 10.1103/PhysRevD.98.115020} {\bibfield  {journal} {\bibinfo
  {journal} {Phys. Rev. D}\ }\textbf {\bibinfo {volume} {98}},\ \bibinfo
  {pages} {115020} (\bibinfo {year} {2018})},\ \Eprint
  {http://arxiv.org/abs/1807.03788} {arXiv:1807.03788 [hep-ph]} \BibitemShut
  {NoStop}%
\bibitem [{\citenamefont {Bodmer}(1971)}]{Bodmer:1971we}%
  \BibitemOpen
  \bibfield  {author} {\bibinfo {author} {\bibfnamefont {A.~R.}\ \bibnamefont
  {Bodmer}},\ }\href {\doibase 10.1103/PhysRevD.4.1601} {\bibfield  {journal}
  {\bibinfo  {journal} {Phys. Rev. D}\ }\textbf {\bibinfo {volume} {4}},\
  \bibinfo {pages} {1601} (\bibinfo {year} {1971})}\BibitemShut {NoStop}%
\bibitem [{\citenamefont {Witten}(1984)}]{Witten:1984rs}%
  \BibitemOpen
  \bibfield  {author} {\bibinfo {author} {\bibfnamefont {E.}~\bibnamefont
  {Witten}},\ }\href {\doibase 10.1103/PhysRevD.30.272} {\bibfield  {journal}
  {\bibinfo  {journal} {Phys. Rev. D}\ }\textbf {\bibinfo {volume} {30}},\
  \bibinfo {pages} {272} (\bibinfo {year} {1984})}\BibitemShut {NoStop}%
\bibitem [{\citenamefont {Bai}\ \emph {et~al.}(2019)\citenamefont {Bai},
  \citenamefont {Long},\ and\ \citenamefont {Lu}}]{Bai:2018dxf}%
  \BibitemOpen
  \bibfield  {author} {\bibinfo {author} {\bibfnamefont {Y.}~\bibnamefont
  {Bai}}, \bibinfo {author} {\bibfnamefont {A.~J.}\ \bibnamefont {Long}}, \
  and\ \bibinfo {author} {\bibfnamefont {S.}~\bibnamefont {Lu}},\ }\href
  {\doibase 10.1103/PhysRevD.99.055047} {\bibfield  {journal} {\bibinfo
  {journal} {Phys. Rev. D}\ }\textbf {\bibinfo {volume} {99}},\ \bibinfo
  {pages} {055047} (\bibinfo {year} {2019})},\ \Eprint
  {http://arxiv.org/abs/1810.04360} {arXiv:1810.04360 [hep-ph]} \BibitemShut
  {NoStop}%
\bibitem [{\citenamefont {Hardy}\ \emph
  {et~al.}(2015{\natexlab{a}})\citenamefont {Hardy}, \citenamefont {Lasenby},
  \citenamefont {March-Russell},\ and\ \citenamefont {West}}]{Hardy:2014mqa}%
  \BibitemOpen
  \bibfield  {author} {\bibinfo {author} {\bibfnamefont {E.}~\bibnamefont
  {Hardy}}, \bibinfo {author} {\bibfnamefont {R.}~\bibnamefont {Lasenby}},
  \bibinfo {author} {\bibfnamefont {J.}~\bibnamefont {March-Russell}}, \ and\
  \bibinfo {author} {\bibfnamefont {S.~M.}\ \bibnamefont {West}},\ }\href
  {\doibase 10.1007/JHEP06(2015)011} {\bibfield  {journal} {\bibinfo  {journal}
  {JHEP}\ }\textbf {\bibinfo {volume} {06}},\ \bibinfo {pages} {011} (\bibinfo
  {year} {2015}{\natexlab{a}})},\ \Eprint {http://arxiv.org/abs/1411.3739}
  {arXiv:1411.3739 [hep-ph]} \BibitemShut {NoStop}%
\bibitem [{\citenamefont {Hardy}\ \emph
  {et~al.}(2015{\natexlab{b}})\citenamefont {Hardy}, \citenamefont {Lasenby},
  \citenamefont {March-Russell},\ and\ \citenamefont {West}}]{Hardy:2015boa}%
  \BibitemOpen
  \bibfield  {author} {\bibinfo {author} {\bibfnamefont {E.}~\bibnamefont
  {Hardy}}, \bibinfo {author} {\bibfnamefont {R.}~\bibnamefont {Lasenby}},
  \bibinfo {author} {\bibfnamefont {J.}~\bibnamefont {March-Russell}}, \ and\
  \bibinfo {author} {\bibfnamefont {S.~M.}\ \bibnamefont {West}},\ }\href
  {\doibase 10.1007/JHEP07(2015)133} {\bibfield  {journal} {\bibinfo  {journal}
  {JHEP}\ }\textbf {\bibinfo {volume} {07}},\ \bibinfo {pages} {133} (\bibinfo
  {year} {2015}{\natexlab{b}})},\ \Eprint {http://arxiv.org/abs/1504.05419}
  {arXiv:1504.05419 [hep-ph]} \BibitemShut {NoStop}%
\bibitem [{\citenamefont {Gresham}\ \emph {et~al.}(2017)\citenamefont
  {Gresham}, \citenamefont {Lou},\ and\ \citenamefont
  {Zurek}}]{Gresham:2017zqi}%
  \BibitemOpen
  \bibfield  {author} {\bibinfo {author} {\bibfnamefont {M.~I.}\ \bibnamefont
  {Gresham}}, \bibinfo {author} {\bibfnamefont {H.~K.}\ \bibnamefont {Lou}}, \
  and\ \bibinfo {author} {\bibfnamefont {K.~M.}\ \bibnamefont {Zurek}},\ }\href
  {\doibase 10.1103/PhysRevD.96.096012} {\bibfield  {journal} {\bibinfo
  {journal} {Phys. Rev. D}\ }\textbf {\bibinfo {volume} {96}},\ \bibinfo
  {pages} {096012} (\bibinfo {year} {2017})},\ \Eprint
  {http://arxiv.org/abs/1707.02313} {arXiv:1707.02313 [hep-ph]} \BibitemShut
  {NoStop}%
\bibitem [{\citenamefont {Redi}\ and\ \citenamefont
  {Tesi}(2019)}]{Redi:2018muu}%
  \BibitemOpen
  \bibfield  {author} {\bibinfo {author} {\bibfnamefont {M.}~\bibnamefont
  {Redi}}\ and\ \bibinfo {author} {\bibfnamefont {A.}~\bibnamefont {Tesi}},\
  }\href {\doibase 10.1007/JHEP04(2019)108} {\bibfield  {journal} {\bibinfo
  {journal} {JHEP}\ }\textbf {\bibinfo {volume} {04}},\ \bibinfo {pages} {108}
  (\bibinfo {year} {2019})},\ \Eprint {http://arxiv.org/abs/1812.08784}
  {arXiv:1812.08784 [hep-ph]} \BibitemShut {NoStop}%
\bibitem [{\citenamefont {Detmold}\ \emph
  {et~al.}(2014{\natexlab{a}})\citenamefont {Detmold}, \citenamefont
  {McCullough},\ and\ \citenamefont {Pochinsky}}]{Detmold:2014qqa}%
  \BibitemOpen
  \bibfield  {author} {\bibinfo {author} {\bibfnamefont {W.}~\bibnamefont
  {Detmold}}, \bibinfo {author} {\bibfnamefont {M.}~\bibnamefont {McCullough}},
  \ and\ \bibinfo {author} {\bibfnamefont {A.}~\bibnamefont {Pochinsky}},\
  }\href {\doibase 10.1103/PhysRevD.90.115013} {\bibfield  {journal} {\bibinfo
  {journal} {Phys. Rev. D}\ }\textbf {\bibinfo {volume} {90}},\ \bibinfo
  {pages} {115013} (\bibinfo {year} {2014}{\natexlab{a}})},\ \Eprint
  {http://arxiv.org/abs/1406.2276} {arXiv:1406.2276 [hep-ph]} \BibitemShut
  {NoStop}%
\bibitem [{\citenamefont {Detmold}\ \emph
  {et~al.}(2014{\natexlab{b}})\citenamefont {Detmold}, \citenamefont
  {McCullough},\ and\ \citenamefont {Pochinsky}}]{Detmold:2014kba}%
  \BibitemOpen
  \bibfield  {author} {\bibinfo {author} {\bibfnamefont {W.}~\bibnamefont
  {Detmold}}, \bibinfo {author} {\bibfnamefont {M.}~\bibnamefont {McCullough}},
  \ and\ \bibinfo {author} {\bibfnamefont {A.}~\bibnamefont {Pochinsky}},\
  }\href {\doibase 10.1103/PhysRevD.90.114506} {\bibfield  {journal} {\bibinfo
  {journal} {Phys. Rev. D}\ }\textbf {\bibinfo {volume} {90}},\ \bibinfo
  {pages} {114506} (\bibinfo {year} {2014}{\natexlab{b}})},\ \Eprint
  {http://arxiv.org/abs/1406.4116} {arXiv:1406.4116 [hep-lat]} \BibitemShut
  {NoStop}%
\bibitem [{\citenamefont {Gresham}\ \emph
  {et~al.}(2018{\natexlab{a}})\citenamefont {Gresham}, \citenamefont {Lou},\
  and\ \citenamefont {Zurek}}]{Gresham:2018anj}%
  \BibitemOpen
  \bibfield  {author} {\bibinfo {author} {\bibfnamefont {M.~I.}\ \bibnamefont
  {Gresham}}, \bibinfo {author} {\bibfnamefont {H.~K.}\ \bibnamefont {Lou}}, \
  and\ \bibinfo {author} {\bibfnamefont {K.~M.}\ \bibnamefont {Zurek}},\ }\href
  {\doibase 10.1103/PhysRevD.98.096001} {\bibfield  {journal} {\bibinfo
  {journal} {Phys. Rev. D}\ }\textbf {\bibinfo {volume} {98}},\ \bibinfo
  {pages} {096001} (\bibinfo {year} {2018}{\natexlab{a}})},\ \Eprint
  {http://arxiv.org/abs/1805.04512} {arXiv:1805.04512 [hep-ph]} \BibitemShut
  {NoStop}%
\bibitem [{\citenamefont {Krnjaic}\ and\ \citenamefont
  {Sigurdson}(2015)}]{Krnjaic:2014xza}%
  \BibitemOpen
  \bibfield  {author} {\bibinfo {author} {\bibfnamefont {G.}~\bibnamefont
  {Krnjaic}}\ and\ \bibinfo {author} {\bibfnamefont {K.}~\bibnamefont
  {Sigurdson}},\ }\href {\doibase 10.1016/j.physletb.2015.11.001} {\bibfield
  {journal} {\bibinfo  {journal} {Phys. Lett. B}\ }\textbf {\bibinfo {volume}
  {751}},\ \bibinfo {pages} {464} (\bibinfo {year} {2015})},\ \Eprint
  {http://arxiv.org/abs/1406.1171} {arXiv:1406.1171 [hep-ph]} \BibitemShut
  {NoStop}%
\bibitem [{\citenamefont {Gresham}\ \emph
  {et~al.}(2018{\natexlab{b}})\citenamefont {Gresham}, \citenamefont {Lou},\
  and\ \citenamefont {Zurek}}]{Gresham:2017cvl}%
  \BibitemOpen
  \bibfield  {author} {\bibinfo {author} {\bibfnamefont {M.~I.}\ \bibnamefont
  {Gresham}}, \bibinfo {author} {\bibfnamefont {H.~K.}\ \bibnamefont {Lou}}, \
  and\ \bibinfo {author} {\bibfnamefont {K.~M.}\ \bibnamefont {Zurek}},\ }\href
  {\doibase 10.1103/PhysRevD.97.036003} {\bibfield  {journal} {\bibinfo
  {journal} {Phys. Rev. D}\ }\textbf {\bibinfo {volume} {97}},\ \bibinfo
  {pages} {036003} (\bibinfo {year} {2018}{\natexlab{b}})},\ \Eprint
  {http://arxiv.org/abs/1707.02316} {arXiv:1707.02316 [hep-ph]} \BibitemShut
  {NoStop}%
\bibitem [{\citenamefont {Coleman}(1985)}]{Coleman:1985ki}%
  \BibitemOpen
  \bibfield  {author} {\bibinfo {author} {\bibfnamefont {S.~R.}\ \bibnamefont
  {Coleman}},\ }\href {\doibase 10.1016/0550-3213(86)90520-1} {\bibfield
  {journal} {\bibinfo  {journal} {Nucl. Phys. B}\ }\textbf {\bibinfo {volume}
  {262}},\ \bibinfo {pages} {263} (\bibinfo {year} {1985})},\ \bibinfo {note}
  {[Erratum: Nucl.Phys.B 269, 744 (1986)]}\BibitemShut {NoStop}%
\bibitem [{\citenamefont {Friedberg}\ \emph {et~al.}(1976)\citenamefont
  {Friedberg}, \citenamefont {Lee},\ and\ \citenamefont
  {Sirlin}}]{Friedberg:1976me}%
  \BibitemOpen
  \bibfield  {author} {\bibinfo {author} {\bibfnamefont {R.}~\bibnamefont
  {Friedberg}}, \bibinfo {author} {\bibfnamefont {T.~D.}\ \bibnamefont {Lee}},
  \ and\ \bibinfo {author} {\bibfnamefont {A.}~\bibnamefont {Sirlin}},\ }\href
  {\doibase 10.1103/PhysRevD.13.2739} {\bibfield  {journal} {\bibinfo
  {journal} {Phys. Rev. D}\ }\textbf {\bibinfo {volume} {13}},\ \bibinfo
  {pages} {2739} (\bibinfo {year} {1976})}\BibitemShut {NoStop}%
\bibitem [{\citenamefont {Kusenko}\ and\ \citenamefont
  {Shaposhnikov}(1998)}]{Kusenko:1997si}%
  \BibitemOpen
  \bibfield  {author} {\bibinfo {author} {\bibfnamefont {A.}~\bibnamefont
  {Kusenko}}\ and\ \bibinfo {author} {\bibfnamefont {M.~E.}\ \bibnamefont
  {Shaposhnikov}},\ }\href {\doibase 10.1016/S0370-2693(97)01375-0} {\bibfield
  {journal} {\bibinfo  {journal} {Phys. Lett. B}\ }\textbf {\bibinfo {volume}
  {418}},\ \bibinfo {pages} {46} (\bibinfo {year} {1998})},\ \Eprint
  {http://arxiv.org/abs/hep-ph/9709492} {arXiv:hep-ph/9709492} \BibitemShut
  {NoStop}%
\bibitem [{\citenamefont {Kusenko}(1997)}]{Kusenko:1997zq}%
  \BibitemOpen
  \bibfield  {author} {\bibinfo {author} {\bibfnamefont {A.}~\bibnamefont
  {Kusenko}},\ }\href {\doibase 10.1016/S0370-2693(97)00584-4} {\bibfield
  {journal} {\bibinfo  {journal} {Phys. Lett. B}\ }\textbf {\bibinfo {volume}
  {405}},\ \bibinfo {pages} {108} (\bibinfo {year} {1997})},\ \Eprint
  {http://arxiv.org/abs/hep-ph/9704273} {arXiv:hep-ph/9704273} \BibitemShut
  {NoStop}%
\bibitem [{\citenamefont {Kusenko}(2006)}]{Kusenko:2006gv}%
  \BibitemOpen
  \bibfield  {author} {\bibinfo {author} {\bibfnamefont {A.}~\bibnamefont
  {Kusenko}},\ }in\ \href@noop {} {\emph {\bibinfo {booktitle} {{Workshop on
  Exotic Physics with Neutrino Telescopes}}}}\ (\bibinfo {year} {2006})\
  \Eprint {http://arxiv.org/abs/hep-ph/0612159} {arXiv:hep-ph/0612159}
  \BibitemShut {NoStop}%
\bibitem [{\citenamefont {Hong}\ \emph {et~al.}(2016)\citenamefont {Hong},
  \citenamefont {Kawasaki},\ and\ \citenamefont {Yamada}}]{Hong:2016ict}%
  \BibitemOpen
  \bibfield  {author} {\bibinfo {author} {\bibfnamefont {J.-P.}\ \bibnamefont
  {Hong}}, \bibinfo {author} {\bibfnamefont {M.}~\bibnamefont {Kawasaki}}, \
  and\ \bibinfo {author} {\bibfnamefont {M.}~\bibnamefont {Yamada}},\ }\href
  {\doibase 10.1088/1475-7516/2016/08/053} {\bibfield  {journal} {\bibinfo
  {journal} {JCAP}\ }\textbf {\bibinfo {volume} {08}},\ \bibinfo {pages} {053}
  (\bibinfo {year} {2016})},\ \Eprint {http://arxiv.org/abs/1604.04352}
  {arXiv:1604.04352 [hep-ph]} \BibitemShut {NoStop}%
\bibitem [{\citenamefont {Hong}\ \emph {et~al.}(2020)\citenamefont {Hong},
  \citenamefont {Jung},\ and\ \citenamefont {Xie}}]{Hong:2020est}%
  \BibitemOpen
  \bibfield  {author} {\bibinfo {author} {\bibfnamefont {J.-P.}\ \bibnamefont
  {Hong}}, \bibinfo {author} {\bibfnamefont {S.}~\bibnamefont {Jung}}, \ and\
  \bibinfo {author} {\bibfnamefont {K.-P.}\ \bibnamefont {Xie}},\ }\href
  {\doibase 10.1103/PhysRevD.102.075028} {\bibfield  {journal} {\bibinfo
  {journal} {Phys. Rev. D}\ }\textbf {\bibinfo {volume} {102}},\ \bibinfo
  {pages} {075028} (\bibinfo {year} {2020})},\ \Eprint
  {http://arxiv.org/abs/2008.04430} {arXiv:2008.04430 [hep-ph]} \BibitemShut
  {NoStop}%
\bibitem [{\citenamefont {Holdom}(1987{\natexlab{a}})}]{Holdom:1987ep}%
  \BibitemOpen
  \bibfield  {author} {\bibinfo {author} {\bibfnamefont {B.}~\bibnamefont
  {Holdom}},\ }\href {\doibase 10.1103/PhysRevD.36.1000} {\bibfield  {journal}
  {\bibinfo  {journal} {Phys. Rev. D}\ }\textbf {\bibinfo {volume} {36}},\
  \bibinfo {pages} {1000} (\bibinfo {year} {1987}{\natexlab{a}})}\BibitemShut
  {NoStop}%
\bibitem [{\citenamefont {Lohiya}(1994)}]{Lohiya:1994pmf}%
  \BibitemOpen
  \bibfield  {author} {\bibinfo {author} {\bibfnamefont {D.}~\bibnamefont
  {Lohiya}},\ }\href@noop {} {\  (\bibinfo {year} {1994})},\ \Eprint
  {http://arxiv.org/abs/gr-qc/9407014} {arXiv:gr-qc/9407014} \BibitemShut
  {NoStop}%
\bibitem [{\citenamefont {Stojkovic}(2003)}]{Stojkovic:2001qi}%
  \BibitemOpen
  \bibfield  {author} {\bibinfo {author} {\bibfnamefont {D.}~\bibnamefont
  {Stojkovic}},\ }\href {\doibase 10.1103/PhysRevD.67.045012} {\bibfield
  {journal} {\bibinfo  {journal} {Phys. Rev. D}\ }\textbf {\bibinfo {volume}
  {67}},\ \bibinfo {pages} {045012} (\bibinfo {year} {2003})},\ \Eprint
  {http://arxiv.org/abs/hep-ph/0111061} {arXiv:hep-ph/0111061} \BibitemShut
  {NoStop}%
\bibitem [{\citenamefont {Macpherson}\ and\ \citenamefont
  {Campbell}(1995)}]{Macpherson:1994wf}%
  \BibitemOpen
  \bibfield  {author} {\bibinfo {author} {\bibfnamefont {A.~L.}\ \bibnamefont
  {Macpherson}}\ and\ \bibinfo {author} {\bibfnamefont {B.~A.}\ \bibnamefont
  {Campbell}},\ }\href {\doibase 10.1016/0370-2693(95)00080-5} {\bibfield
  {journal} {\bibinfo  {journal} {Phys. Lett. B}\ }\textbf {\bibinfo {volume}
  {347}},\ \bibinfo {pages} {205} (\bibinfo {year} {1995})},\ \Eprint
  {http://arxiv.org/abs/hep-ph/9408387} {arXiv:hep-ph/9408387} \BibitemShut
  {NoStop}%
\bibitem [{\citenamefont {Holdom}(1987{\natexlab{b}})}]{Holdom:1987bu}%
  \BibitemOpen
  \bibfield  {author} {\bibinfo {author} {\bibfnamefont {B.}~\bibnamefont
  {Holdom}},\ }\href@noop {} {\  (\bibinfo {year}
  {1987}{\natexlab{b}})}\BibitemShut {NoStop}%
\bibitem [{\citenamefont {Zhitnitsky}(2003)}]{Zhitnitsky:2002qa}%
  \BibitemOpen
  \bibfield  {author} {\bibinfo {author} {\bibfnamefont {A.~R.}\ \bibnamefont
  {Zhitnitsky}},\ }\href {\doibase 10.1088/1475-7516/2003/10/010} {\bibfield
  {journal} {\bibinfo  {journal} {JCAP}\ }\textbf {\bibinfo {volume} {10}},\
  \bibinfo {pages} {010} (\bibinfo {year} {2003})},\ \Eprint
  {http://arxiv.org/abs/hep-ph/0202161} {arXiv:hep-ph/0202161} \BibitemShut
  {NoStop}%
\bibitem [{\citenamefont {Ogure}\ \emph {et~al.}(2003)\citenamefont {Ogure},
  \citenamefont {Yoshida},\ and\ \citenamefont {Arafune}}]{Ogure:2002hv}%
  \BibitemOpen
  \bibfield  {author} {\bibinfo {author} {\bibfnamefont {K.}~\bibnamefont
  {Ogure}}, \bibinfo {author} {\bibfnamefont {T.}~\bibnamefont {Yoshida}}, \
  and\ \bibinfo {author} {\bibfnamefont {J.}~\bibnamefont {Arafune}},\ }\href
  {\doibase 10.1103/PhysRevD.67.123518} {\bibfield  {journal} {\bibinfo
  {journal} {Phys. Rev. D}\ }\textbf {\bibinfo {volume} {67}},\ \bibinfo
  {pages} {123518} (\bibinfo {year} {2003})},\ \Eprint
  {http://arxiv.org/abs/hep-ph/0212332} {arXiv:hep-ph/0212332} \BibitemShut
  {NoStop}%
\bibitem [{\citenamefont {Frieman}\ \emph {et~al.}(1988)\citenamefont
  {Frieman}, \citenamefont {Gelmini}, \citenamefont {Gleiser},\ and\
  \citenamefont {Kolb}}]{Frieman:1988ut}%
  \BibitemOpen
  \bibfield  {author} {\bibinfo {author} {\bibfnamefont {J.~A.}\ \bibnamefont
  {Frieman}}, \bibinfo {author} {\bibfnamefont {G.~B.}\ \bibnamefont
  {Gelmini}}, \bibinfo {author} {\bibfnamefont {M.}~\bibnamefont {Gleiser}}, \
  and\ \bibinfo {author} {\bibfnamefont {E.~W.}\ \bibnamefont {Kolb}},\ }\href
  {\doibase 10.1103/PhysRevLett.60.2101} {\bibfield  {journal} {\bibinfo
  {journal} {Phys. Rev. Lett.}\ }\textbf {\bibinfo {volume} {60}},\ \bibinfo
  {pages} {2101} (\bibinfo {year} {1988})}\BibitemShut {NoStop}%
\bibitem [{\citenamefont {Affleck}\ and\ \citenamefont
  {Dine}(1985)}]{Affleck:1984fy}%
  \BibitemOpen
  \bibfield  {author} {\bibinfo {author} {\bibfnamefont {I.}~\bibnamefont
  {Affleck}}\ and\ \bibinfo {author} {\bibfnamefont {M.}~\bibnamefont {Dine}},\
  }\href {\doibase 10.1016/0550-3213(85)90021-5} {\bibfield  {journal}
  {\bibinfo  {journal} {Nucl. Phys. B}\ }\textbf {\bibinfo {volume} {249}},\
  \bibinfo {pages} {361} (\bibinfo {year} {1985})}\BibitemShut {NoStop}%
\bibitem [{\citenamefont {Dine}\ and\ \citenamefont
  {Kusenko}(2003)}]{Dine:2003ax}%
  \BibitemOpen
  \bibfield  {author} {\bibinfo {author} {\bibfnamefont {M.}~\bibnamefont
  {Dine}}\ and\ \bibinfo {author} {\bibfnamefont {A.}~\bibnamefont {Kusenko}},\
  }\href {\doibase 10.1103/RevModPhys.76.1} {\bibfield  {journal} {\bibinfo
  {journal} {Rev. Mod. Phys.}\ }\textbf {\bibinfo {volume} {76}},\ \bibinfo
  {pages} {1} (\bibinfo {year} {2003})},\ \Eprint
  {http://arxiv.org/abs/hep-ph/0303065} {arXiv:hep-ph/0303065} \BibitemShut
  {NoStop}%
\bibitem [{\citenamefont {Krylov}\ \emph {et~al.}(2013)\citenamefont {Krylov},
  \citenamefont {Levin},\ and\ \citenamefont {Rubakov}}]{Krylov:2013qe}%
  \BibitemOpen
  \bibfield  {author} {\bibinfo {author} {\bibfnamefont {E.}~\bibnamefont
  {Krylov}}, \bibinfo {author} {\bibfnamefont {A.}~\bibnamefont {Levin}}, \
  and\ \bibinfo {author} {\bibfnamefont {V.}~\bibnamefont {Rubakov}},\ }\href
  {\doibase 10.1103/PhysRevD.87.083528} {\bibfield  {journal} {\bibinfo
  {journal} {Phys. Rev. D}\ }\textbf {\bibinfo {volume} {87}},\ \bibinfo
  {pages} {083528} (\bibinfo {year} {2013})},\ \Eprint
  {http://arxiv.org/abs/1301.0354} {arXiv:1301.0354 [hep-ph]} \BibitemShut
  {NoStop}%
\bibitem [{\citenamefont {Chung}\ \emph {et~al.}(2001)\citenamefont {Chung},
  \citenamefont {Crotty}, \citenamefont {Kolb},\ and\ \citenamefont
  {Riotto}}]{chung_crotty_kolb_riotto_2001}%
  \BibitemOpen
  \bibfield  {author} {\bibinfo {author} {\bibfnamefont {D.~J.~H.}\
  \bibnamefont {Chung}}, \bibinfo {author} {\bibfnamefont {P.}~\bibnamefont
  {Crotty}}, \bibinfo {author} {\bibfnamefont {E.~W.}\ \bibnamefont {Kolb}}, \
  and\ \bibinfo {author} {\bibfnamefont {A.}~\bibnamefont {Riotto}},\ }\href
  {\doibase 10.1103/physrevd.64.043503} {\bibfield  {journal} {\bibinfo
  {journal} {Physical Review D}\ }\textbf {\bibinfo {volume} {64}} (\bibinfo
  {year} {2001}),\ 10.1103/physrevd.64.043503}\BibitemShut {NoStop}%
\bibitem [{\citenamefont {Chung}\ \emph
  {et~al.}(1998{\natexlab{a}})\citenamefont {Chung}, \citenamefont {Kolb},\
  and\ \citenamefont {Riotto}}]{Chung:1998zb}%
  \BibitemOpen
  \bibfield  {author} {\bibinfo {author} {\bibfnamefont {D.~J.~H.}\
  \bibnamefont {Chung}}, \bibinfo {author} {\bibfnamefont {E.~W.}\ \bibnamefont
  {Kolb}}, \ and\ \bibinfo {author} {\bibfnamefont {A.}~\bibnamefont
  {Riotto}},\ }\href {\doibase 10.1103/PhysRevD.59.023501} {\bibfield
  {journal} {\bibinfo  {journal} {Phys. Rev. D}\ }\textbf {\bibinfo {volume}
  {59}},\ \bibinfo {pages} {023501} (\bibinfo {year} {1998}{\natexlab{a}})},\
  \Eprint {http://arxiv.org/abs/hep-ph/9802238} {arXiv:hep-ph/9802238}
  \BibitemShut {NoStop}%
\bibitem [{\citenamefont {Chung}\ \emph
  {et~al.}(1998{\natexlab{b}})\citenamefont {Chung}, \citenamefont {Kolb},\
  and\ \citenamefont {Riotto}}]{Chung:1998ua}%
  \BibitemOpen
  \bibfield  {author} {\bibinfo {author} {\bibfnamefont {D.~J.~H.}\
  \bibnamefont {Chung}}, \bibinfo {author} {\bibfnamefont {E.~W.}\ \bibnamefont
  {Kolb}}, \ and\ \bibinfo {author} {\bibfnamefont {A.}~\bibnamefont
  {Riotto}},\ }\href {\doibase 10.1103/PhysRevLett.81.4048} {\bibfield
  {journal} {\bibinfo  {journal} {Phys. Rev. Lett.}\ }\textbf {\bibinfo
  {volume} {81}},\ \bibinfo {pages} {4048} (\bibinfo {year}
  {1998}{\natexlab{b}})},\ \Eprint {http://arxiv.org/abs/hep-ph/9805473}
  {arXiv:hep-ph/9805473} \BibitemShut {NoStop}%
\bibitem [{\citenamefont {Hooper}\ \emph {et~al.}(2019)\citenamefont {Hooper},
  \citenamefont {Krnjaic},\ and\ \citenamefont {McDermott}}]{Hooper:2019gtx}%
  \BibitemOpen
  \bibfield  {author} {\bibinfo {author} {\bibfnamefont {D.}~\bibnamefont
  {Hooper}}, \bibinfo {author} {\bibfnamefont {G.}~\bibnamefont {Krnjaic}}, \
  and\ \bibinfo {author} {\bibfnamefont {S.~D.}\ \bibnamefont {McDermott}},\
  }\href {\doibase 10.1007/JHEP08(2019)001} {\bibfield  {journal} {\bibinfo
  {journal} {JHEP}\ }\textbf {\bibinfo {volume} {08}},\ \bibinfo {pages} {001}
  (\bibinfo {year} {2019})},\ \Eprint {http://arxiv.org/abs/1905.01301}
  {arXiv:1905.01301 [hep-ph]} \BibitemShut {NoStop}%
\bibitem [{\citenamefont {Bernal}\ and\ \citenamefont
  {Zapata}(2021)}]{Bernal:2020bjf}%
  \BibitemOpen
  \bibfield  {author} {\bibinfo {author} {\bibfnamefont {N.}~\bibnamefont
  {Bernal}}\ and\ \bibinfo {author} {\bibfnamefont {O.}~\bibnamefont
  {Zapata}},\ }\href {\doibase 10.1088/1475-7516/2021/03/015} {\bibfield
  {journal} {\bibinfo  {journal} {JCAP}\ }\textbf {\bibinfo {volume} {03}},\
  \bibinfo {pages} {015} (\bibinfo {year} {2021})},\ \Eprint
  {http://arxiv.org/abs/2011.12306} {arXiv:2011.12306 [astro-ph.CO]}
  \BibitemShut {NoStop}%
\bibitem [{\citenamefont {Lehmann}\ \emph {et~al.}(2019)\citenamefont
  {Lehmann}, \citenamefont {Johnson}, \citenamefont {Profumo},\ and\
  \citenamefont {Schwemberger}}]{lehmann_johnson_profumo_schwemberger_2019}%
  \BibitemOpen
  \bibfield  {author} {\bibinfo {author} {\bibfnamefont {B.~V.}\ \bibnamefont
  {Lehmann}}, \bibinfo {author} {\bibfnamefont {C.}~\bibnamefont {Johnson}},
  \bibinfo {author} {\bibfnamefont {S.}~\bibnamefont {Profumo}}, \ and\
  \bibinfo {author} {\bibfnamefont {T.}~\bibnamefont {Schwemberger}},\ }\href
  {\doibase 10.1088/1475-7516/2019/10/046} {\bibfield  {journal} {\bibinfo
  {journal} {Journal of Cosmology and Astroparticle Physics}\ }\textbf
  {\bibinfo {volume} {2019}},\ \bibinfo {pages} {046–046} (\bibinfo {year}
  {2019})}\BibitemShut {NoStop}%
\bibitem [{\citenamefont {Bai}\ and\ \citenamefont
  {Orlofsky}(2020)}]{Bai:2019zcd}%
  \BibitemOpen
  \bibfield  {author} {\bibinfo {author} {\bibfnamefont {Y.}~\bibnamefont
  {Bai}}\ and\ \bibinfo {author} {\bibfnamefont {N.}~\bibnamefont {Orlofsky}},\
  }\href {\doibase 10.1103/PhysRevD.101.055006} {\bibfield  {journal} {\bibinfo
   {journal} {Phys. Rev. D}\ }\textbf {\bibinfo {volume} {101}},\ \bibinfo
  {pages} {055006} (\bibinfo {year} {2020})},\ \Eprint
  {http://arxiv.org/abs/1906.04858} {arXiv:1906.04858 [hep-ph]} \BibitemShut
  {NoStop}%
\bibitem [{\citenamefont {Aydemir}\ \emph {et~al.}(2020)\citenamefont
  {Aydemir}, \citenamefont {Holdom},\ and\ \citenamefont
  {Ren}}]{Aydemir:2020xfd}%
  \BibitemOpen
  \bibfield  {author} {\bibinfo {author} {\bibfnamefont {U.}~\bibnamefont
  {Aydemir}}, \bibinfo {author} {\bibfnamefont {B.}~\bibnamefont {Holdom}}, \
  and\ \bibinfo {author} {\bibfnamefont {J.}~\bibnamefont {Ren}},\ }\href
  {\doibase 10.1103/PhysRevD.102.024058} {\bibfield  {journal} {\bibinfo
  {journal} {Phys. Rev. D}\ }\textbf {\bibinfo {volume} {102}},\ \bibinfo
  {pages} {024058} (\bibinfo {year} {2020})},\ \Eprint
  {http://arxiv.org/abs/2003.10682} {arXiv:2003.10682 [gr-qc]} \BibitemShut
  {NoStop}%
\bibitem [{\citenamefont {MacGibbon}(1987)}]{MacGibbon:1987my}%
  \BibitemOpen
  \bibfield  {author} {\bibinfo {author} {\bibfnamefont {J.~H.}\ \bibnamefont
  {MacGibbon}},\ }\href {\doibase 10.1038/329308a0} {\bibfield  {journal}
  {\bibinfo  {journal} {Nature}\ }\textbf {\bibinfo {volume} {329}},\ \bibinfo
  {pages} {308} (\bibinfo {year} {1987})}\BibitemShut {NoStop}%
\bibitem [{\citenamefont {Salvio}\ and\ \citenamefont
  {Veerm\"ae}(2020)}]{Salvio:2019llz}%
  \BibitemOpen
  \bibfield  {author} {\bibinfo {author} {\bibfnamefont {A.}~\bibnamefont
  {Salvio}}\ and\ \bibinfo {author} {\bibfnamefont {H.}~\bibnamefont
  {Veerm\"ae}},\ }\href {\doibase 10.1088/1475-7516/2020/02/018} {\bibfield
  {journal} {\bibinfo  {journal} {JCAP}\ }\textbf {\bibinfo {volume} {02}},\
  \bibinfo {pages} {018} (\bibinfo {year} {2020})},\ \Eprint
  {http://arxiv.org/abs/1912.13333} {arXiv:1912.13333 [gr-qc]} \BibitemShut
  {NoStop}%
\bibitem [{\citenamefont {Agnese}\ \emph {et~al.}(2014)\citenamefont {Agnese},
  \citenamefont {Anderson}, \citenamefont {Asai}, \citenamefont
  {Balakishiyeva}, \citenamefont {Thakur}, \citenamefont {Bauer}, \citenamefont
  {Beaty}, \citenamefont {Billard}, \citenamefont {Borgland}, \citenamefont
  {Bowles} \emph {et~al.}}]{agnese2014search}%
  \BibitemOpen
  \bibfield  {author} {\bibinfo {author} {\bibfnamefont {R.}~\bibnamefont
  {Agnese}}, \bibinfo {author} {\bibfnamefont {A.~J.}\ \bibnamefont
  {Anderson}}, \bibinfo {author} {\bibfnamefont {M.}~\bibnamefont {Asai}},
  \bibinfo {author} {\bibfnamefont {D.}~\bibnamefont {Balakishiyeva}}, \bibinfo
  {author} {\bibfnamefont {R.~B.}\ \bibnamefont {Thakur}}, \bibinfo {author}
  {\bibfnamefont {D.}~\bibnamefont {Bauer}}, \bibinfo {author} {\bibfnamefont
  {J.}~\bibnamefont {Beaty}}, \bibinfo {author} {\bibfnamefont
  {J.}~\bibnamefont {Billard}}, \bibinfo {author} {\bibfnamefont
  {A.}~\bibnamefont {Borgland}}, \bibinfo {author} {\bibfnamefont
  {M.}~\bibnamefont {Bowles}},  \emph {et~al.},\ }\href@noop {} {\bibfield
  {journal} {\bibinfo  {journal} {Physical review letters}\ }\textbf {\bibinfo
  {volume} {112}},\ \bibinfo {pages} {241302} (\bibinfo {year}
  {2014})}\BibitemShut {NoStop}%
\bibitem [{\citenamefont {Lawrie}\ \emph {et~al.}(2021)\citenamefont {Lawrie},
  \citenamefont {Feldman}, \citenamefont {Marvinney},\ and\ \citenamefont
  {Pai}}]{lawrie2021free}%
  \BibitemOpen
  \bibfield  {author} {\bibinfo {author} {\bibfnamefont {B.~J.}\ \bibnamefont
  {Lawrie}}, \bibinfo {author} {\bibfnamefont {M.}~\bibnamefont {Feldman}},
  \bibinfo {author} {\bibfnamefont {C.~E.}\ \bibnamefont {Marvinney}}, \ and\
  \bibinfo {author} {\bibfnamefont {Y.-Y.}\ \bibnamefont {Pai}},\ }in\
  \href@noop {} {\emph {\bibinfo {booktitle} {Quantum Nanophotonic Materials,
  Devices, and Systems 2021}}},\ Vol.\ \bibinfo {volume} {11806}\ (\bibinfo
  {organization} {International Society for Optics and Photonics},\ \bibinfo
  {year} {2021})\ p.\ \bibinfo {pages} {1180604}\BibitemShut {NoStop}%
\bibitem [{\citenamefont {Gr{\"o}bner}(2001)}]{grobner2001overview}%
  \BibitemOpen
  \bibfield  {author} {\bibinfo {author} {\bibfnamefont {O.}~\bibnamefont
  {Gr{\"o}bner}},\ }\href@noop {} {\bibfield  {journal} {\bibinfo  {journal}
  {Vacuum}\ }\textbf {\bibinfo {volume} {60}},\ \bibinfo {pages} {25} (\bibinfo
  {year} {2001})}\BibitemShut {NoStop}%
\bibitem [{\citenamefont {Caves}(1981)}]{caves1981quantum}%
  \BibitemOpen
  \bibfield  {author} {\bibinfo {author} {\bibfnamefont {C.~M.}\ \bibnamefont
  {Caves}},\ }\href@noop {} {\bibfield  {journal} {\bibinfo  {journal}
  {Physical Review D}\ }\textbf {\bibinfo {volume} {23}},\ \bibinfo {pages}
  {1693} (\bibinfo {year} {1981})}\BibitemShut {NoStop}%
\bibitem [{\citenamefont {Aasi}\ \emph {et~al.}(2013)\citenamefont {Aasi},
  \citenamefont {Abadie}, \citenamefont {Abbott}, \citenamefont {Abbott},
  \citenamefont {Abbott}, \citenamefont {Abernathy}, \citenamefont {Adams},
  \citenamefont {Adams}, \citenamefont {Addesso}, \citenamefont {Adhikari}
  \emph {et~al.}}]{aasi2013enhanced}%
  \BibitemOpen
  \bibfield  {author} {\bibinfo {author} {\bibfnamefont {J.}~\bibnamefont
  {Aasi}}, \bibinfo {author} {\bibfnamefont {J.}~\bibnamefont {Abadie}},
  \bibinfo {author} {\bibfnamefont {B.}~\bibnamefont {Abbott}}, \bibinfo
  {author} {\bibfnamefont {R.}~\bibnamefont {Abbott}}, \bibinfo {author}
  {\bibfnamefont {T.}~\bibnamefont {Abbott}}, \bibinfo {author} {\bibfnamefont
  {M.}~\bibnamefont {Abernathy}}, \bibinfo {author} {\bibfnamefont
  {C.}~\bibnamefont {Adams}}, \bibinfo {author} {\bibfnamefont
  {T.}~\bibnamefont {Adams}}, \bibinfo {author} {\bibfnamefont
  {P.}~\bibnamefont {Addesso}}, \bibinfo {author} {\bibfnamefont
  {R.}~\bibnamefont {Adhikari}},  \emph {et~al.},\ }\href@noop {} {\bibfield
  {journal} {\bibinfo  {journal} {Nature Photonics}\ }\textbf {\bibinfo
  {volume} {7}},\ \bibinfo {pages} {613} (\bibinfo {year} {2013})}\BibitemShut
  {NoStop}%
\bibitem [{\citenamefont {Vahlbruch}\ \emph
  {et~al.}(2016{\natexlab{a}})\citenamefont {Vahlbruch}, \citenamefont
  {Mehmet}, \citenamefont {Danzmann},\ and\ \citenamefont
  {Schnabel}}]{PhysRevLett.117.110801}%
  \BibitemOpen
  \bibfield  {author} {\bibinfo {author} {\bibfnamefont {H.}~\bibnamefont
  {Vahlbruch}}, \bibinfo {author} {\bibfnamefont {M.}~\bibnamefont {Mehmet}},
  \bibinfo {author} {\bibfnamefont {K.}~\bibnamefont {Danzmann}}, \ and\
  \bibinfo {author} {\bibfnamefont {R.}~\bibnamefont {Schnabel}},\ }\href
  {\doibase 10.1103/PhysRevLett.117.110801} {\bibfield  {journal} {\bibinfo
  {journal} {Phys. Rev. Lett.}\ }\textbf {\bibinfo {volume} {117}},\ \bibinfo
  {pages} {110801} (\bibinfo {year} {2016}{\natexlab{a}})}\BibitemShut
  {NoStop}%
\bibitem [{\citenamefont {Caves}\ \emph {et~al.}(1980)\citenamefont {Caves},
  \citenamefont {Thorne}, \citenamefont {Drever}, \citenamefont {Sandberg},\
  and\ \citenamefont {Zimmermann}}]{caves1980measurement}%
  \BibitemOpen
  \bibfield  {author} {\bibinfo {author} {\bibfnamefont {C.~M.}\ \bibnamefont
  {Caves}}, \bibinfo {author} {\bibfnamefont {K.~S.}\ \bibnamefont {Thorne}},
  \bibinfo {author} {\bibfnamefont {R.~W.}\ \bibnamefont {Drever}}, \bibinfo
  {author} {\bibfnamefont {V.~D.}\ \bibnamefont {Sandberg}}, \ and\ \bibinfo
  {author} {\bibfnamefont {M.}~\bibnamefont {Zimmermann}},\ }\href@noop {}
  {\bibfield  {journal} {\bibinfo  {journal} {Reviews of Modern Physics}\
  }\textbf {\bibinfo {volume} {52}},\ \bibinfo {pages} {341} (\bibinfo {year}
  {1980})}\BibitemShut {NoStop}%
\bibitem [{\citenamefont {Braginsky}\ and\ \citenamefont
  {Khalili}(1990)}]{braginsky1990gravitational}%
  \BibitemOpen
  \bibfield  {author} {\bibinfo {author} {\bibfnamefont {V.~B.}\ \bibnamefont
  {Braginsky}}\ and\ \bibinfo {author} {\bibfnamefont {F.~J.}\ \bibnamefont
  {Khalili}},\ }\href@noop {} {\bibfield  {journal} {\bibinfo  {journal}
  {Physics Letters A}\ }\textbf {\bibinfo {volume} {147}},\ \bibinfo {pages}
  {251} (\bibinfo {year} {1990})}\BibitemShut {NoStop}%
\bibitem [{\citenamefont {Ghosh}\ \emph {et~al.}(2019)\citenamefont {Ghosh},
  \citenamefont {Carney}, \citenamefont {Shawhan},\ and\ \citenamefont
  {Taylor}}]{Ghosh:2019rsc}%
  \BibitemOpen
  \bibfield  {author} {\bibinfo {author} {\bibfnamefont {S.}~\bibnamefont
  {Ghosh}}, \bibinfo {author} {\bibfnamefont {D.}~\bibnamefont {Carney}},
  \bibinfo {author} {\bibfnamefont {P.}~\bibnamefont {Shawhan}}, \ and\
  \bibinfo {author} {\bibfnamefont {J.~M.}\ \bibnamefont {Taylor}},\
  }\href@noop {} {\  (\bibinfo {year} {2019})},\ \Eprint
  {http://arxiv.org/abs/1910.11892} {arXiv:1910.11892 [quant-ph]} \BibitemShut
  {NoStop}%
\bibitem [{\citenamefont {Glorieux}\ \emph {et~al.}(2010)\citenamefont
  {Glorieux}, \citenamefont {Guidoni}, \citenamefont {Guibal}, \citenamefont
  {Likforman},\ and\ \citenamefont {Coudreau}}]{glorieux2010strong}%
  \BibitemOpen
  \bibfield  {author} {\bibinfo {author} {\bibfnamefont {Q.}~\bibnamefont
  {Glorieux}}, \bibinfo {author} {\bibfnamefont {L.}~\bibnamefont {Guidoni}},
  \bibinfo {author} {\bibfnamefont {S.}~\bibnamefont {Guibal}}, \bibinfo
  {author} {\bibfnamefont {J.-P.}\ \bibnamefont {Likforman}}, \ and\ \bibinfo
  {author} {\bibfnamefont {T.}~\bibnamefont {Coudreau}},\ }in\ \href@noop {}
  {\emph {\bibinfo {booktitle} {Quantum Optics}}},\ Vol.\ \bibinfo {volume}
  {7727}\ (\bibinfo {organization} {SPIE},\ \bibinfo {year} {2010})\ pp.\
  \bibinfo {pages} {9--16}\BibitemShut {NoStop}%
\bibitem [{\citenamefont {Vahlbruch}\ \emph
  {et~al.}(2016{\natexlab{b}})\citenamefont {Vahlbruch}, \citenamefont
  {Mehmet}, \citenamefont {Danzmann},\ and\ \citenamefont
  {Schnabel}}]{vahlbruch2016detection}%
  \BibitemOpen
  \bibfield  {author} {\bibinfo {author} {\bibfnamefont {H.}~\bibnamefont
  {Vahlbruch}}, \bibinfo {author} {\bibfnamefont {M.}~\bibnamefont {Mehmet}},
  \bibinfo {author} {\bibfnamefont {K.}~\bibnamefont {Danzmann}}, \ and\
  \bibinfo {author} {\bibfnamefont {R.}~\bibnamefont {Schnabel}},\ }\href@noop
  {} {\bibfield  {journal} {\bibinfo  {journal} {Physical review letters}\
  }\textbf {\bibinfo {volume} {117}},\ \bibinfo {pages} {110801} (\bibinfo
  {year} {2016}{\natexlab{b}})}\BibitemShut {NoStop}%
\bibitem [{\citenamefont {Lawrie}\ \emph {et~al.}(2019)\citenamefont {Lawrie},
  \citenamefont {Lett}, \citenamefont {Marino},\ and\ \citenamefont
  {Pooser}}]{lawrie2019quantum}%
  \BibitemOpen
  \bibfield  {author} {\bibinfo {author} {\bibfnamefont {B.~J.}\ \bibnamefont
  {Lawrie}}, \bibinfo {author} {\bibfnamefont {P.~D.}\ \bibnamefont {Lett}},
  \bibinfo {author} {\bibfnamefont {A.~M.}\ \bibnamefont {Marino}}, \ and\
  \bibinfo {author} {\bibfnamefont {R.~C.}\ \bibnamefont {Pooser}},\
  }\href@noop {} {\bibfield  {journal} {\bibinfo  {journal} {ACS Photonics}\
  }\textbf {\bibinfo {volume} {6}},\ \bibinfo {pages} {1307} (\bibinfo {year}
  {2019})}\BibitemShut {NoStop}%
\bibitem [{\citenamefont {Anderson}\ \emph {et~al.}(2017)\citenamefont
  {Anderson}, \citenamefont {Gupta}, \citenamefont {Schmittberger},
  \citenamefont {Horrom}, \citenamefont {Hermann-Avigliano}, \citenamefont
  {Jones},\ and\ \citenamefont {Lett}}]{anderson2017phase}%
  \BibitemOpen
  \bibfield  {author} {\bibinfo {author} {\bibfnamefont {B.~E.}\ \bibnamefont
  {Anderson}}, \bibinfo {author} {\bibfnamefont {P.}~\bibnamefont {Gupta}},
  \bibinfo {author} {\bibfnamefont {B.~L.}\ \bibnamefont {Schmittberger}},
  \bibinfo {author} {\bibfnamefont {T.}~\bibnamefont {Horrom}}, \bibinfo
  {author} {\bibfnamefont {C.}~\bibnamefont {Hermann-Avigliano}}, \bibinfo
  {author} {\bibfnamefont {K.~M.}\ \bibnamefont {Jones}}, \ and\ \bibinfo
  {author} {\bibfnamefont {P.~D.}\ \bibnamefont {Lett}},\ }\href@noop {}
  {\bibfield  {journal} {\bibinfo  {journal} {Optica}\ }\textbf {\bibinfo
  {volume} {4}},\ \bibinfo {pages} {752} (\bibinfo {year} {2017})}\BibitemShut
  {NoStop}%
\bibitem [{\citenamefont {Pooser}\ \emph {et~al.}(2020)\citenamefont {Pooser},
  \citenamefont {Savino}, \citenamefont {Batson}, \citenamefont {Beckey},
  \citenamefont {Garcia},\ and\ \citenamefont {Lawrie}}]{pooser2020truncated}%
  \BibitemOpen
  \bibfield  {author} {\bibinfo {author} {\bibfnamefont {R.}~\bibnamefont
  {Pooser}}, \bibinfo {author} {\bibfnamefont {N.}~\bibnamefont {Savino}},
  \bibinfo {author} {\bibfnamefont {E.}~\bibnamefont {Batson}}, \bibinfo
  {author} {\bibfnamefont {J.}~\bibnamefont {Beckey}}, \bibinfo {author}
  {\bibfnamefont {J.}~\bibnamefont {Garcia}}, \ and\ \bibinfo {author}
  {\bibfnamefont {B.}~\bibnamefont {Lawrie}},\ }\href@noop {} {\bibfield
  {journal} {\bibinfo  {journal} {Physical Review Letters}\ }\textbf {\bibinfo
  {volume} {124}},\ \bibinfo {pages} {230504} (\bibinfo {year}
  {2020})}\BibitemShut {NoStop}%
\bibitem [{\citenamefont {Pooser}\ and\ \citenamefont
  {Lawrie}(2015)}]{pooser2015ultrasensitive}%
  \BibitemOpen
  \bibfield  {author} {\bibinfo {author} {\bibfnamefont {R.~C.}\ \bibnamefont
  {Pooser}}\ and\ \bibinfo {author} {\bibfnamefont {B.}~\bibnamefont
  {Lawrie}},\ }\href@noop {} {\bibfield  {journal} {\bibinfo  {journal}
  {Optica}\ }\textbf {\bibinfo {volume} {2}},\ \bibinfo {pages} {393} (\bibinfo
  {year} {2015})}\BibitemShut {NoStop}%
\bibitem [{\citenamefont {Pai}\ \emph {et~al.}(2022)\citenamefont {Pai},
  \citenamefont {Marvinney}, \citenamefont {Hua}, \citenamefont {Pooser},\ and\
  \citenamefont {Lawrie}}]{pai2022magneto}%
  \BibitemOpen
  \bibfield  {author} {\bibinfo {author} {\bibfnamefont {Y.-Y.}\ \bibnamefont
  {Pai}}, \bibinfo {author} {\bibfnamefont {C.~E.}\ \bibnamefont {Marvinney}},
  \bibinfo {author} {\bibfnamefont {C.}~\bibnamefont {Hua}}, \bibinfo {author}
  {\bibfnamefont {R.~C.}\ \bibnamefont {Pooser}}, \ and\ \bibinfo {author}
  {\bibfnamefont {B.~J.}\ \bibnamefont {Lawrie}},\ }\href@noop {} {\bibfield
  {journal} {\bibinfo  {journal} {Advanced Quantum Technologies}\ }\textbf
  {\bibinfo {volume} {5}},\ \bibinfo {pages} {2100107} (\bibinfo {year}
  {2022})}\BibitemShut {NoStop}%
\bibitem [{\citenamefont {Zhou}\ \emph {et~al.}(2021)\citenamefont {Zhou},
  \citenamefont {Bao}, \citenamefont {Madugani}, \citenamefont {Long},
  \citenamefont {Gorman},\ and\ \citenamefont {LeBrun}}]{zhou2021}%
  \BibitemOpen
  \bibfield  {author} {\bibinfo {author} {\bibfnamefont {F.}~\bibnamefont
  {Zhou}}, \bibinfo {author} {\bibfnamefont {Y.}~\bibnamefont {Bao}}, \bibinfo
  {author} {\bibfnamefont {R.}~\bibnamefont {Madugani}}, \bibinfo {author}
  {\bibfnamefont {D.~A.}\ \bibnamefont {Long}}, \bibinfo {author}
  {\bibfnamefont {J.~J.}\ \bibnamefont {Gorman}}, \ and\ \bibinfo {author}
  {\bibfnamefont {T.~W.}\ \bibnamefont {LeBrun}},\ }\href@noop {} {\bibfield
  {journal} {\bibinfo  {journal} {Optica}\ }\textbf {\bibinfo {volume} {8}},\
  \bibinfo {pages} {350} (\bibinfo {year} {2021})}\BibitemShut {NoStop}%
\bibitem [{\citenamefont {Guzm{\'a}n~Cervantes}\ \emph
  {et~al.}(2014)\citenamefont {Guzm{\'a}n~Cervantes}, \citenamefont
  {Kumanchik}, \citenamefont {Pratt},\ and\ \citenamefont
  {Taylor}}]{guzman2014}%
  \BibitemOpen
  \bibfield  {author} {\bibinfo {author} {\bibfnamefont {F.}~\bibnamefont
  {Guzm{\'a}n~Cervantes}}, \bibinfo {author} {\bibfnamefont {L.}~\bibnamefont
  {Kumanchik}}, \bibinfo {author} {\bibfnamefont {J.}~\bibnamefont {Pratt}}, \
  and\ \bibinfo {author} {\bibfnamefont {J.~M.}\ \bibnamefont {Taylor}},\
  }\href@noop {} {\bibfield  {journal} {\bibinfo  {journal} {Applied Physics
  Letters}\ }\textbf {\bibinfo {volume} {104}},\ \bibinfo {pages} {221111}
  (\bibinfo {year} {2014})}\BibitemShut {NoStop}%
\bibitem [{\citenamefont {Hutchison}\ and\ \citenamefont
  {Bhave}(2012)}]{hutchison2012z}%
  \BibitemOpen
  \bibfield  {author} {\bibinfo {author} {\bibfnamefont {D.~N.}\ \bibnamefont
  {Hutchison}}\ and\ \bibinfo {author} {\bibfnamefont {S.~A.}\ \bibnamefont
  {Bhave}},\ }in\ \href@noop {} {\emph {\bibinfo {booktitle} {2012 IEEE 25th
  international conference on micro electro mechanical systems (MEMS)}}}\
  (\bibinfo {organization} {IEEE},\ \bibinfo {year} {2012})\ pp.\ \bibinfo
  {pages} {615--619}\BibitemShut {NoStop}%
\bibitem [{\citenamefont {Tian}\ \emph {et~al.}(2020)\citenamefont {Tian},
  \citenamefont {Liu}, \citenamefont {Dong}, \citenamefont {Skehan},
  \citenamefont {Zervas}, \citenamefont {Kippenberg},\ and\ \citenamefont
  {Bhave}}]{tian2020}%
  \BibitemOpen
  \bibfield  {author} {\bibinfo {author} {\bibfnamefont {H.}~\bibnamefont
  {Tian}}, \bibinfo {author} {\bibfnamefont {J.}~\bibnamefont {Liu}}, \bibinfo
  {author} {\bibfnamefont {B.}~\bibnamefont {Dong}}, \bibinfo {author}
  {\bibfnamefont {J.~C.}\ \bibnamefont {Skehan}}, \bibinfo {author}
  {\bibfnamefont {M.}~\bibnamefont {Zervas}}, \bibinfo {author} {\bibfnamefont
  {T.~J.}\ \bibnamefont {Kippenberg}}, \ and\ \bibinfo {author} {\bibfnamefont
  {S.~A.}\ \bibnamefont {Bhave}},\ }\href@noop {} {\bibfield  {journal}
  {\bibinfo  {journal} {Nature communications}\ }\textbf {\bibinfo {volume}
  {11}},\ \bibinfo {pages} {1} (\bibinfo {year} {2020})}\BibitemShut {NoStop}%
\bibitem [{\citenamefont {Liu}\ \emph {et~al.}(2020)\citenamefont {Liu},
  \citenamefont {Tian}, \citenamefont {Lucas}, \citenamefont {Raja},
  \citenamefont {Lihachev}, \citenamefont {Wang}, \citenamefont {He},
  \citenamefont {Liu}, \citenamefont {Anderson}, \citenamefont {Weng} \emph
  {et~al.}}]{liu2020}%
  \BibitemOpen
  \bibfield  {author} {\bibinfo {author} {\bibfnamefont {J.}~\bibnamefont
  {Liu}}, \bibinfo {author} {\bibfnamefont {H.}~\bibnamefont {Tian}}, \bibinfo
  {author} {\bibfnamefont {E.}~\bibnamefont {Lucas}}, \bibinfo {author}
  {\bibfnamefont {A.~S.}\ \bibnamefont {Raja}}, \bibinfo {author}
  {\bibfnamefont {G.}~\bibnamefont {Lihachev}}, \bibinfo {author}
  {\bibfnamefont {R.~N.}\ \bibnamefont {Wang}}, \bibinfo {author}
  {\bibfnamefont {J.}~\bibnamefont {He}}, \bibinfo {author} {\bibfnamefont
  {T.}~\bibnamefont {Liu}}, \bibinfo {author} {\bibfnamefont {M.~H.}\
  \bibnamefont {Anderson}}, \bibinfo {author} {\bibfnamefont {W.}~\bibnamefont
  {Weng}},  \emph {et~al.},\ }\href@noop {} {\bibfield  {journal} {\bibinfo
  {journal} {Nature}\ }\textbf {\bibinfo {volume} {583}},\ \bibinfo {pages}
  {385} (\bibinfo {year} {2020})}\BibitemShut {NoStop}%
\bibitem [{\citenamefont {Krause}\ \emph {et~al.}(2012)\citenamefont {Krause},
  \citenamefont {Winger}, \citenamefont {Blasius}, \citenamefont {Lin},\ and\
  \citenamefont {Painter}}]{krause2012}%
  \BibitemOpen
  \bibfield  {author} {\bibinfo {author} {\bibfnamefont {A.~G.}\ \bibnamefont
  {Krause}}, \bibinfo {author} {\bibfnamefont {M.}~\bibnamefont {Winger}},
  \bibinfo {author} {\bibfnamefont {T.~D.}\ \bibnamefont {Blasius}}, \bibinfo
  {author} {\bibfnamefont {Q.}~\bibnamefont {Lin}}, \ and\ \bibinfo {author}
  {\bibfnamefont {O.}~\bibnamefont {Painter}},\ }\href@noop {} {\bibfield
  {journal} {\bibinfo  {journal} {Nature Photonics}\ }\textbf {\bibinfo
  {volume} {6}},\ \bibinfo {pages} {768} (\bibinfo {year} {2012})}\BibitemShut
  {NoStop}%
\bibitem [{\citenamefont {Zhang}\ \emph {et~al.}(2019)\citenamefont {Zhang},
  \citenamefont {Buscaino}, \citenamefont {Wang}, \citenamefont {Shams-Ansari},
  \citenamefont {Reimer}, \citenamefont {Zhu}, \citenamefont {Kahn},\ and\
  \citenamefont {Lon{\v{c}}ar}}]{zhang2019}%
  \BibitemOpen
  \bibfield  {author} {\bibinfo {author} {\bibfnamefont {M.}~\bibnamefont
  {Zhang}}, \bibinfo {author} {\bibfnamefont {B.}~\bibnamefont {Buscaino}},
  \bibinfo {author} {\bibfnamefont {C.}~\bibnamefont {Wang}}, \bibinfo {author}
  {\bibfnamefont {A.}~\bibnamefont {Shams-Ansari}}, \bibinfo {author}
  {\bibfnamefont {C.}~\bibnamefont {Reimer}}, \bibinfo {author} {\bibfnamefont
  {R.}~\bibnamefont {Zhu}}, \bibinfo {author} {\bibfnamefont {J.~M.}\
  \bibnamefont {Kahn}}, \ and\ \bibinfo {author} {\bibfnamefont
  {M.}~\bibnamefont {Lon{\v{c}}ar}},\ }\href@noop {} {\bibfield  {journal}
  {\bibinfo  {journal} {Nature}\ }\textbf {\bibinfo {volume} {568}},\ \bibinfo
  {pages} {373} (\bibinfo {year} {2019})}\BibitemShut {NoStop}%
\bibitem [{\citenamefont {Foudas}\ \emph {et~al.}(2005)\citenamefont {Foudas}
  \emph {et~al.}}]{Foudas:2004axy}%
  \BibitemOpen
  \bibfield  {author} {\bibinfo {author} {\bibfnamefont {C.}~\bibnamefont
  {Foudas}} \emph {et~al.},\ }\href {\doibase 10.1109/TNS.2005.860173}
  {\bibfield  {journal} {\bibinfo  {journal} {IEEE Trans. Nucl. Sci.}\ }\textbf
  {\bibinfo {volume} {52}},\ \bibinfo {pages} {2836} (\bibinfo {year}
  {2005})},\ \Eprint {http://arxiv.org/abs/physics/0510229}
  {arXiv:physics/0510229} \BibitemShut {NoStop}%
\bibitem [{\citenamefont {{Buchner}}(2021)}]{2021JOSS....6.3001B}%
  \BibitemOpen
  \bibfield  {author} {\bibinfo {author} {\bibfnamefont {J.}~\bibnamefont
  {{Buchner}}},\ }\href {\doibase 10.21105/joss.03001} {\bibfield  {journal}
  {\bibinfo  {journal} {The Journal of Open Source Software}\ }\textbf
  {\bibinfo {volume} {6}},\ \bibinfo {eid} {3001} (\bibinfo {year} {2021})},\
  \Eprint {http://arxiv.org/abs/2101.09604} {arXiv:2101.09604 [stat.CO]}
  \BibitemShut {NoStop}%
\bibitem [{\citenamefont {Chaudhuri}\ \emph {et~al.}(2018)\citenamefont
  {Chaudhuri}, \citenamefont {Irwin}, \citenamefont {Graham},\ and\
  \citenamefont {Mardon}}]{Chaudhuri:2018rqn}%
  \BibitemOpen
  \bibfield  {author} {\bibinfo {author} {\bibfnamefont {S.}~\bibnamefont
  {Chaudhuri}}, \bibinfo {author} {\bibfnamefont {K.}~\bibnamefont {Irwin}},
  \bibinfo {author} {\bibfnamefont {P.~W.}\ \bibnamefont {Graham}}, \ and\
  \bibinfo {author} {\bibfnamefont {J.}~\bibnamefont {Mardon}},\ }\href@noop {}
  {\  (\bibinfo {year} {2018})},\ \Eprint {http://arxiv.org/abs/1803.01627}
  {arXiv:1803.01627 [hep-ph]} \BibitemShut {NoStop}%
\bibitem [{\citenamefont {Salemi}\ \emph {et~al.}(2021)\citenamefont {Salemi},
  \citenamefont {Foster}, \citenamefont {Ouellet}, \citenamefont {Gavin},
  \citenamefont {Pappas} \emph {et~al.}}]{Salemi:2021}%
  \BibitemOpen
  \bibfield  {author} {\bibinfo {author} {\bibfnamefont {C.~P.}\ \bibnamefont
  {Salemi}}, \bibinfo {author} {\bibfnamefont {J.~W.}\ \bibnamefont {Foster}},
  \bibinfo {author} {\bibfnamefont {J.~L.}\ \bibnamefont {Ouellet}}, \bibinfo
  {author} {\bibfnamefont {A.}~\bibnamefont {Gavin}}, \bibinfo {author}
  {\bibfnamefont {K.~M.}\ \bibnamefont {Pappas}},  \emph {et~al.},\ }\href@noop
  {} {\bibfield  {journal} {\bibinfo  {journal} {Physical Review Letters}\
  }\textbf {\bibinfo {volume} {127}} (\bibinfo {year} {2021})}\BibitemShut
  {NoStop}%
\end{thebibliography}%

\end{document}